\documentclass{iopart}

\usepackage{graphicx} 
\usepackage[usenames]{color} 
\usepackage{subfigure} 
\usepackage[colorlinks=true]{hyperref} 
\usepackage{psfrag}  
\usepackage[numbers,square,comma,sort&compress]{natbib} 
\usepackage{iopams}  
\usepackage[normalem]{ulem} 

\definecolor{solarRed}{RGB}{220,50,47} 
\definecolor{orange}{RGB}{220,140,10} 
\definecolor{purple}{RGB}{100,0,160} 
\definecolor{dkred}{RGB}{160,60,50} 

\newcommand{\vect}[1]{\underline{\smash{#1}}}

\newcommand{\eqn}[1]{\eref{#1}}

\newcommand{\fig}[1]{\fref{#1}}
\newcommand{\figs}[1]{figures~\ref{#1}}
\newcommand{\Fig}[1]{\Fref{#1}}
\newcommand{\Figs}[1]{Figures~\ref{#1}}

\newcommand{\tab}[1]{\tref{#1}}
\newcommand{\sect}[1]{\sref{#1}}
\newcommand{\Ref}[1]{Ref.~\cite{#1}}
\newcommand{\brac}[1]{\ensuremath{\left(#1\right)}}
\newcommand{\order}[1]{\ensuremath{\Or\brac{#1}}}
\newcommand{\codevar}[1]{\texttt{#1}}

\newcommand{\reftxt}{\rm ref}
\newcommand{\sa}{s-\ensuremath{\alpha}}
\newcommand{\apar}{\ensuremath{\textrm{A}_{\|}}{}}
\newcommand{\kx}{\ensuremath{k_x}}
\newcommand{\ky}{\ensuremath{k_y}}

\newcommand{\kpar}{\ensuremath{k_\|}}
\newcommand{\wde}{\ensuremath{\omega_{*e}}}
\newcommand{\vthe}{\ensuremath{v_{th,e}}}
\newcommand{\vthi}{\ensuremath{v_{th,i}}}

\newcommand{\vpar}{\ensuremath{v_\|}}
\newcommand{\vperp}{\ensuremath{v_\bot}}

\newcommand{\shat}{\ensuremath{\hat{s}}}

\newcommand{\rhoi}{\ensuremath{\rho_{i}}}
\newcommand{\rhoe}{\ensuremath{\rho_{e}}}

\newcommand{\nue}{\ensuremath{\nu_{ei}}}
\newcommand{\nub}{\ensuremath{\bar{\nu}}}
\newcommand{\lref}{\ensuremath{L_{\reftxt}}}
\newcommand{\lte}{\ensuremath{L_{T_e}}}
\newcommand{\lne}{\ensuremath{L_{n_e}}}
\newcommand{\bfac}{\ensuremath{X}}
\newcommand{\gmtm}{\ensuremath{\gamma_{ \scriptscriptstyle\rm\textrm{\tiny MTM}}}}
\newcommand{\dg}{\ensuremath{\delta g}}
\newcommand{\sgn}{\mathop{\mathrm{sgn}}}

\begin{document}


\title[Microtearing modes at the top of the pedestal]{Microtearing modes at the top of the pedestal}
\author{
D~Dickinson$^1$,
C~M~Roach$^1$,
S~Saarelma$^1$,
R~Scannell$^1$,
A~Kirk$^1$ and
H~R~Wilson$^2$
}

\address{
	$^1$
	EURATOM/CCFE Association,
	Abingdon,
	Oxon,
	OX14 3DB,
	UK.
}
\address{
	$^2$
	York Plasma Institute,
	Department of Physics,
	University of York,
	York,
	YO10 5DD,
	UK
}

\ead{david.dickinson@ccfe.ac.uk}


\begin{abstract}
Microtearing modes (MTMs) are unstable in the shallow gradient region just inside the top of the pedestal in the spherical tokamak experiment MAST, and may play an important role in the pedestal evolution.
The linear properties of these instabilities are compared with MTMs deeper inside the core, and further detailed investigations in \sa{} geometry expose the basic drive mechanism, which is not well described by existing theories.
In particular the growth rate of the dominant edge MTM does not peak at a finite collision frequency, as frequently reported for MTMs further into the core.
Our study suggests that the edge MTM is driven by a collisionless trapped particle mechanism that is sensitive to magnetic drifts.
This drive is enhanced in the outer region of MAST at high magnetic shear and high trapped particle fraction. 
Observations of similar modes in conventional aspect ratio devices suggests this drive mechanism may be somewhat ubiquitous towards the edge of current day and future hot tokamaks.
\end{abstract}

\section{Introduction}
\label{sec:Intro}
Initial analytic studies suggested that tearing modes should be stable at high binormal perpendicular wavenumber, \ky{}, due to increased field line bending \cite{Furth1963}, leading to a focus on larger scale, ``gross'' tearing modes.
A kinetic study of the tearing mode found that an energy dependent collision operator could lead to an additional drive from the electron temperature gradient \cite{Hazeltine1975}.
This drive can overcome the stabilising influence at large $\ky$, allowing unstable microtearing modes (MTMs) to exist.
The parameter $\nub=\nue/\omega$, is important for MTMs, where $\nue$ is the electron-ion collision frequency and $\omega$ is the frequency associated with the mode.
Analytic treatments in the collisional ($\nub\gg1$), semi-collisional ($\nub>1$ and $\nub\ll \kpar^2\vthe^2/\omega^2$) and collisionless ($\nub\ll1$) regimes were performed in slab geometry \cite{Drake1977}.
Here the instability arises due to the time dependent parallel thermal force providing an asymmetry in the parallel force on electrons as a consequence of the energy dependence of the collision operator \cite{Hassam1980,Hassam1980a}.
In the absence of collisions this asymmetry disappears and the slab drive is therefore expected to vanish for sufficiently small $\nub$.
Likewise large $\nub$ becomes stabilising as the collisions prevent the electrons from building a perturbed current.
Kinetic calculations in large aspect ratio toroidal geometry revealed an additional drive mechanism that depends on trapped particles and is effective in the banana regime, $\nub<r/R=\epsilon$ \cite{Catto1981a}.
The trapped particles themselves do not carry the perturbed current but collisions generate the instability by allowing current to grow in the barely passing particles close to the trapped-passing boundary, in a process that remains effective even for $\nub\ll1$.
This effect is found, for realistic tokamak parameters, to be lost at higher collision frequencies in the collisionless regime, $\nub<1<\nub/\epsilon$ \cite{Connor1990}.
Importantly combining the slab and trapped particle drive mechanisms leads to an MTM growth rate which peaks for $\nub\sim\order{1}$.
Observations of a strong inverse collisionality dependence of the thermal confinement time made on both MAST \cite{Valovic2011} and NSTX \cite{Kaye2007a} may be consistent with simulations showing MTM driven transport increasing with $\nue$ \cite{Guttenfelder2012a}.

Fully electromagnetic gyrokinetic simulations are now able to study microtearing modes numerically in experimentally relevant scenarios.
Linear gyrokinetic studies have found unstable MTMs in a wide range of equilibria including at mid-radius in spherical tokamaks (STs)  \cite{Kotschenreuther2000,Applegate2004,Wilson2004a,Roach2005,Applegate2007,Guttenfelder2012}, in simple large aspect ratio shifted circle model equilibria \cite{Applegate2007}, towards the edge in ASDEX Upgrade \cite{Told2008} and during improved confinement in reversed field pinches (RFPs) \cite{Predebon2010,Carmody2012}.
MTMs exhibit tearing parity, where in ballooning space the perturbed parallel magnetic vector potential, \apar{}, is even in the ballooning co-ordinate $\theta$.
This is associated with reconnection at the rational surfaces generating 
small scale island structures.

When the amplitude of these islands is sufficient they will overlap to generate a stochastic field, which gives rise to significant electron heat transport \cite{Stix1973}.
Estimates of the electron thermal diffusivity in NSTX based on a model of stochastic field transport \cite{Rechester1978} are found to be within a factor 2 of the experimental levels over a region in which MTMs are the dominant instability \cite{Wong2007}.
The first successful nonlinear simulations of microtearing turbulence \cite{Doerk2011,Guttenfelder2011} indicate that, in the absence of sheared flows, the associated electron heat flux can indeed be significant, and within the range of experimental observations.
The effect of sheared equilibrium flows is not yet clear, with conflicting findings emerging from these studies.
The dependence of microtearing turbulence on electron beta, $\beta_e=2\mu_0n_eT_e/B^2$, normalised inverse electron temperature gradient scale length $\lref/\lte$ (where  $\lte = T_e/(dT_e/dr)$ and \lref{} is a reference equilibrium length), and collision frequency, $\nue$ \cite{Doerk2012,Guttenfelder2012a} is broadly consistent with previous linear studies \cite{Applegate2007,Told2008,Guttenfelder2012}.
Whilst these gyrokinetic simulations agree qualitatively with the two drive mechanisms discussed earlier, through the dependence of the MTM growth rate, $\gmtm$, on $\nub$, the existence of a critical $dT_e/dr$ and the observation that $\omega\sim\wde$, there is evidence that magnetic drifts, which have not been adequately treated analytically, are also important \cite{Applegate2007}.
In particular, the energy dependence of the collision operator is vital for both analytic drives but in numerical simulations this had little impact in the presence of magnetic drifts \cite{Applegate2007}.
Indeed it was found that both magnetic drifts and the perturbed electrostatic potential, $\phi$, could be destabilising, and in the absence of both of these effects the MTM was found to be stable \cite{Applegate2007}.
It is likely that there are multiple mechanisms occurring simultaneously to drive (or damp) MTMs, with the local parameters determining the relative contribution of each mechanism.
This can lead to different scalings of $\gmtm$ with equilibrium parameters, depending upon which mechanism is dominant.
For example, \Ref{Doerk2012} notes that $\phi$ is destabilising for low safety factor, $q\lesssim3$, but stabilising for $q\gtrsim3$, suggesting that the dominant driving mechanism may be undergoing a transition as $q$ varies.
Whilst two MTM drive mechanisms have been uncovered by analytic theory, it seems that additional mechanisms, involving magnetic drifts, are absent from the existing literature.

\begin{figure}[htb]
    \centering
    \resizebox{0.8\columnwidth}{!}{
    	\subfigure[]{
    	    \psfrag {-100} [cc][cc][0.9][0] {-100}
		    \psfrag {-50} [cc][cc][0.9][0] {-50}
		    \psfrag {0} [cc][cc][0.9][0] {0}
		    \psfrag {50} [cc][cc][0.9][0] {50}
		    \psfrag {100} [cc][cc][0.9][0] {100}
		    \psfrag {q} [ct][cb][1][0] 
		    	{$\theta$}
		    \psfrag {-1.0} [cc][cc][0.9][0] 	
		    	{\hspace{-1em}-1.0}
		    \psfrag {-0.5} [cc][cc][0.9][0]
		    	{\hspace{-1em}-0.5}
		    \psfrag {0.0} [cc][cc][0.9][0] 
		    	{\hspace{-1em}0.0}
		    \psfrag {0.5} [cc][cc][0.9][0] 
		    	{\hspace{-1em}0.5}
		    \psfrag {1.0} [cc][cc][0.9][0] 
		    	{\hspace{-1em}1.0}
    		\psfrag {j} [cc][lc][1][0] 	
    			{\raisebox{3ex}{$\phi$}}
			\label{fig:Intro:CoreEdgePhiComp:Core}
			\includegraphics[width=0.4\columnwidth]
		 	{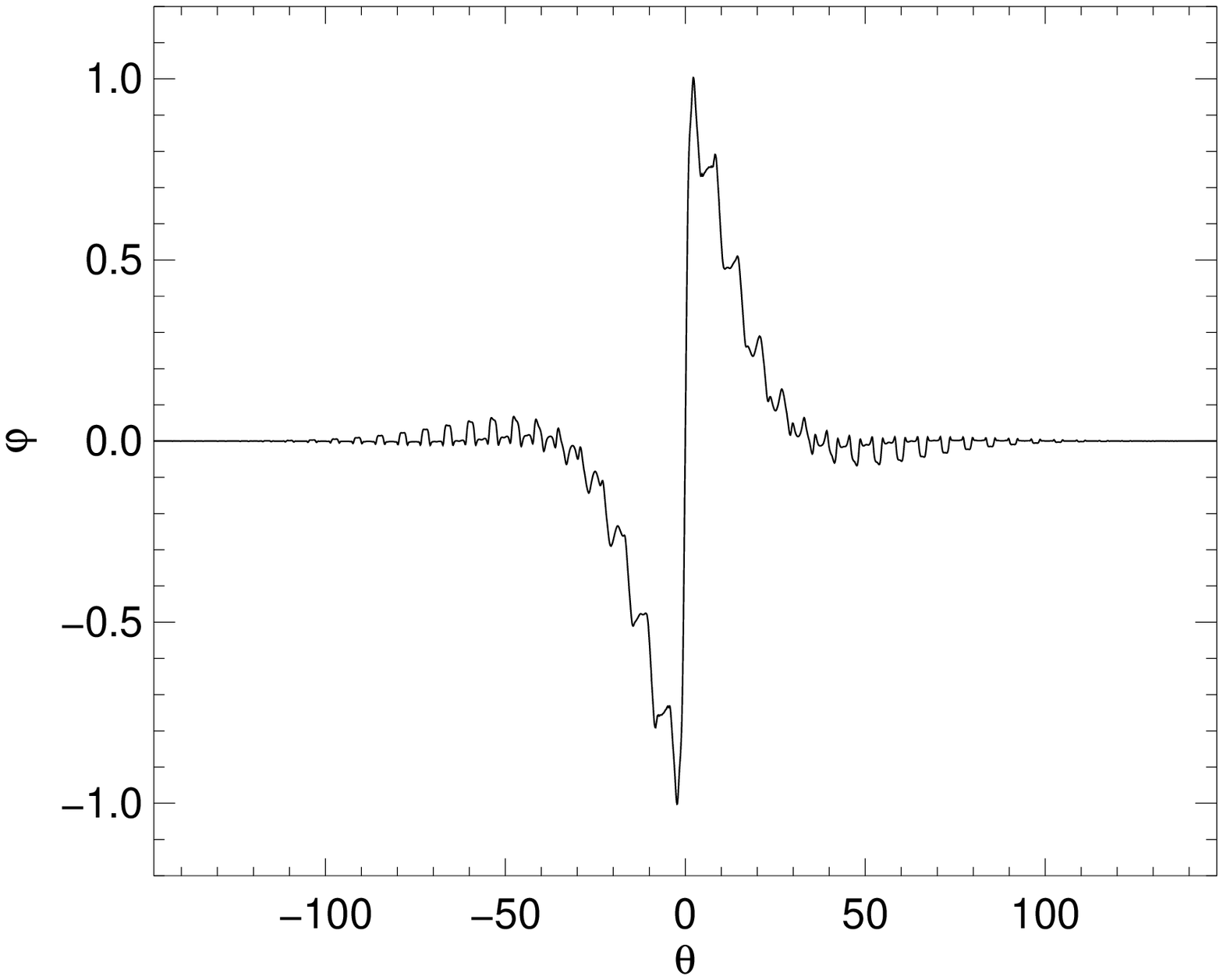}
		}
        \hspace{0.001\columnwidth}
        \subfigure[]{
            \psfrag {q} [ct][cb][1][0] 
            	{$\theta$}
		    \psfrag {-1.0} [cc][cc][0.9][0] 	
		    	{\hspace{-1em}-1.0}
		    \psfrag {-0.5} [cc][cc][0.9][0]
		    	{\hspace{-1em}-0.5}
		    \psfrag {0.0} [cc][cc][0.9][0] 
		    	{\hspace{-1em}0.0}
		    \psfrag {0.5} [cc][cc][0.9][0] 
		    	{\hspace{-1em}0.5}
		    \psfrag {1.0} [cc][cc][0.9][0] 
		    	{\hspace{-1em}1.0}
    		\psfrag {j} [cc][lc][1][0] 
    			{\raisebox{3ex}{$\phi$}}
    		\psfrag {5} [cc][cc][0.9][0] {5}
    		\psfrag {0} [cc][cc][0.9][0] {0}
		    \psfrag {-5} [cc][cc][0.9][0] {-5}			
       	   	\label{fig:Intro:CoreEdgePhiComp:Edge}
        	\includegraphics[width=0.4\columnwidth]
			{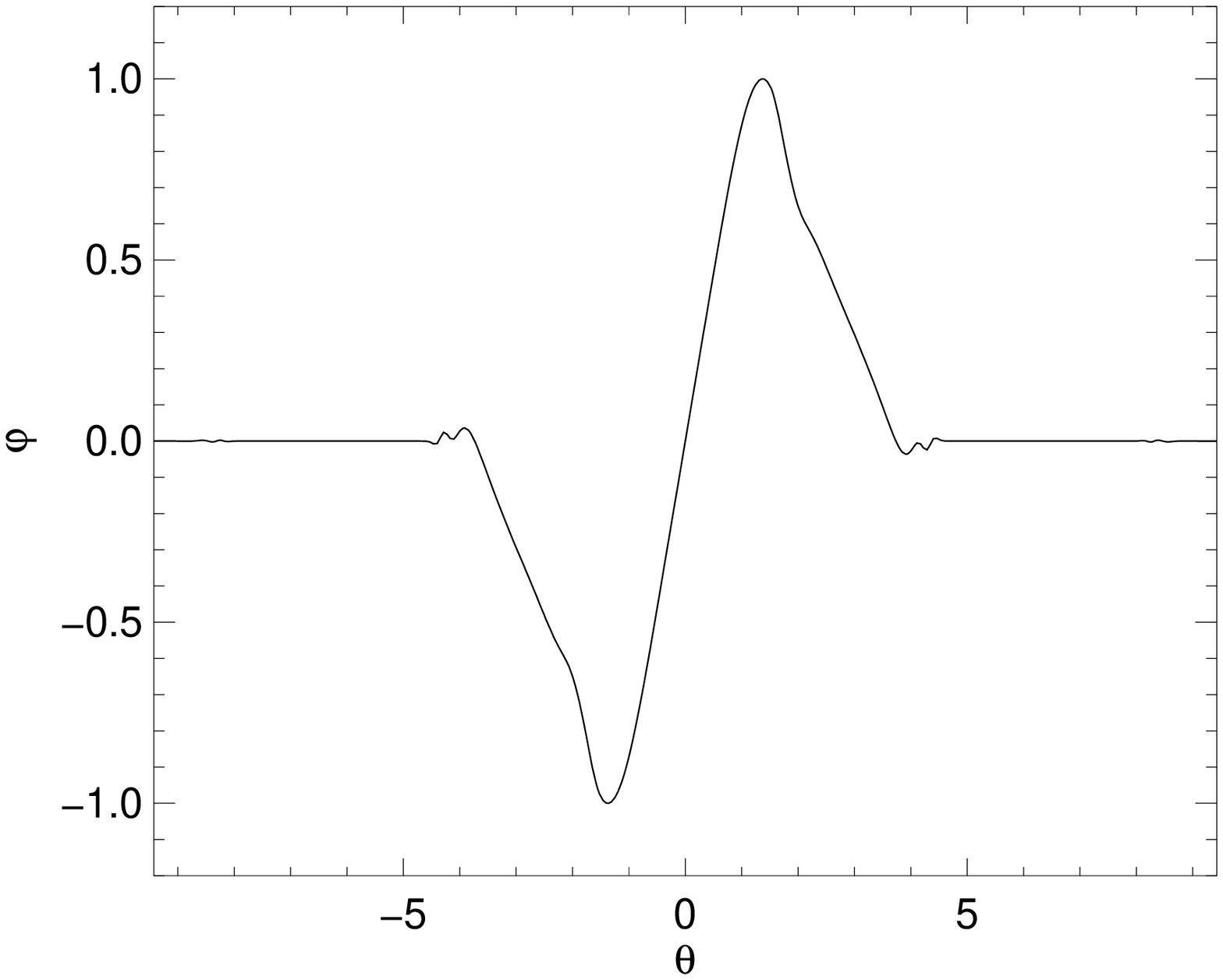}
        }
	}
    \caption[Comparison of core and edge MTM potential fluctuations]{
    The electrostatic potential perturbation, $\phi$, due to MTMs from linear GS2 simulations of MAST at (a) $\psi_N$=0.58 (shot \#27905) and (b) $\psi_N$=0.94 (composite based on shots \#24452, \#24459 and \#24763, see \Ref{Dickinson2011}).
    }
    \label{fig:Intro:CoreEdgePhiComp}
\end{figure}

Recent linear gyrokinetic studies of the edge plasma region in MAST \cite{Dickinson2011} and JET \cite{Saarelma2012} utilising the fully electromagnetic initial value gyrokinetic code GS2 \cite{Kotschenreuther1995} have found unstable MTMs in the shallow gradient region at the top of the pedestal, where they may play an important role in the pedestal evolution \cite{Dickinson2012,RoachIAEA2012}.
This paper provides an in-depth study of such edge MTMs, which whilst related, exhibit significant differences to the more familiar MTMs in the core.
In both edge and core cases the $\apar$ eigenfunctions peak around $\theta=0$ and decay by $\theta=\pm\pi$.
\Fig{fig:Intro:CoreEdgePhiComp}, on the other hand, shows that $\phi$ is considerably less extended in $\theta$ in the edge than in the core, amplifying a similar trend observed in comparisons of $\phi$ from MTMs at $r/a=0.6$ and $r/a=0.8$ in NSTX \cite{Guttenfelder2012}.

%
%

It should be noted that the magnetic shear, $\shat$, is much larger in the edge than in the core, and in both locations $\phi$ extends in $\theta$ to include contributions from large normalised radial wavenumber, $\kx\rhoi=\shat\theta\ky\rhoi\sim\order{10^2}$, where \rhoi{} is the ion Larmor radius.
The radial wavenumber approaches $\kx\delta_0\sim\order{1}$, where the semi-collisional width $\delta_0$ \cite{Drake1977} is defined:
\begin{equation}
\delta_0=
	L_s
	\frac{\sqrt{\wde\nue}}{\ky\vthe}
\label{eq:Intro:SemiCollSkinDepth}
\end{equation}
with the shear length, $L_s=Rq/\shat$.
These cases are both in a similar collisionality regime as $\nub=0.26$ and $0.44$ for the core and edge cases respectively.

In \sect{sec:StudyPar} we introduce a local equilibrium from the MAST edge that is unstable to MTMs, and reduce this to a simpler model equilibrium with similar microstability properties. 
This provides a reference equilibrium for detailed linear gyrokinetic studies, presented in \sect{sec:ModeChar}, that probe the basic driving mechanisms for MTMs in edge plasmas.
Final conclusions are presented in \sect{sec:Conclusion}

\section{Equilibrium parameters and simplifications}
\label{sec:StudyPar}
\begin{figure}[htb]
    \centering
    \resizebox{0.95\columnwidth}{!}{
    	\subfigure[]{
		    \psfrag {0} [cc][cc][1][0] {0}
		    \psfrag {1} [cc][cc][1][0] {1}
		    \psfrag {2} [cc][cc][1][0] {2}
		    \psfrag {3} [cc][cc][1][0] {3}
		    \psfrag {4} [cc][cc][1][0] {4}
		    \psfrag {5} [cc][cc][1][0] {5}
		    \psfrag {6} [cc][cc][1][0] {6}
		    \psfrag {k} [ct][cb][1][0]
		    	{\raisebox{-2.5ex}{\hspace{1em}\ky$\rho_i$}}
		    \psfrag {y} [cc][cc][1][0] {}
		    \psfrag {r} [cc][cc][1][0] {}
		    \psfrag {i} [cc][cc][1][0] {}
		    \psfrag {0.0} [cc][cc][1][0] 
		    	{\hspace{-1em}0.0}
		    \psfrag {0.5} [cc][cc][1][0] 
		    	{\hspace{-1em}0.5}
		    \psfrag {1.0} [cc][cc][1][0] 
		    	{\hspace{-1em}1.0}
		    \psfrag {1.5} [cc][cc][1][0] 
		    	{\hspace{-1em}1.5}
		    \psfrag {g} [cb][lt][1][0] 
		    	{\raisebox{2ex}{\hspace{3em}$\gamma$ ($\vthi/\lref$)}}
		    \psfrag { vthi/Lr} [cc][cc][1][0] {}
			\label{fig:Expt:Spectra:Gamma}
			\raisebox{5ex}{
				\includegraphics[width=0.4\columnwidth]
		 		{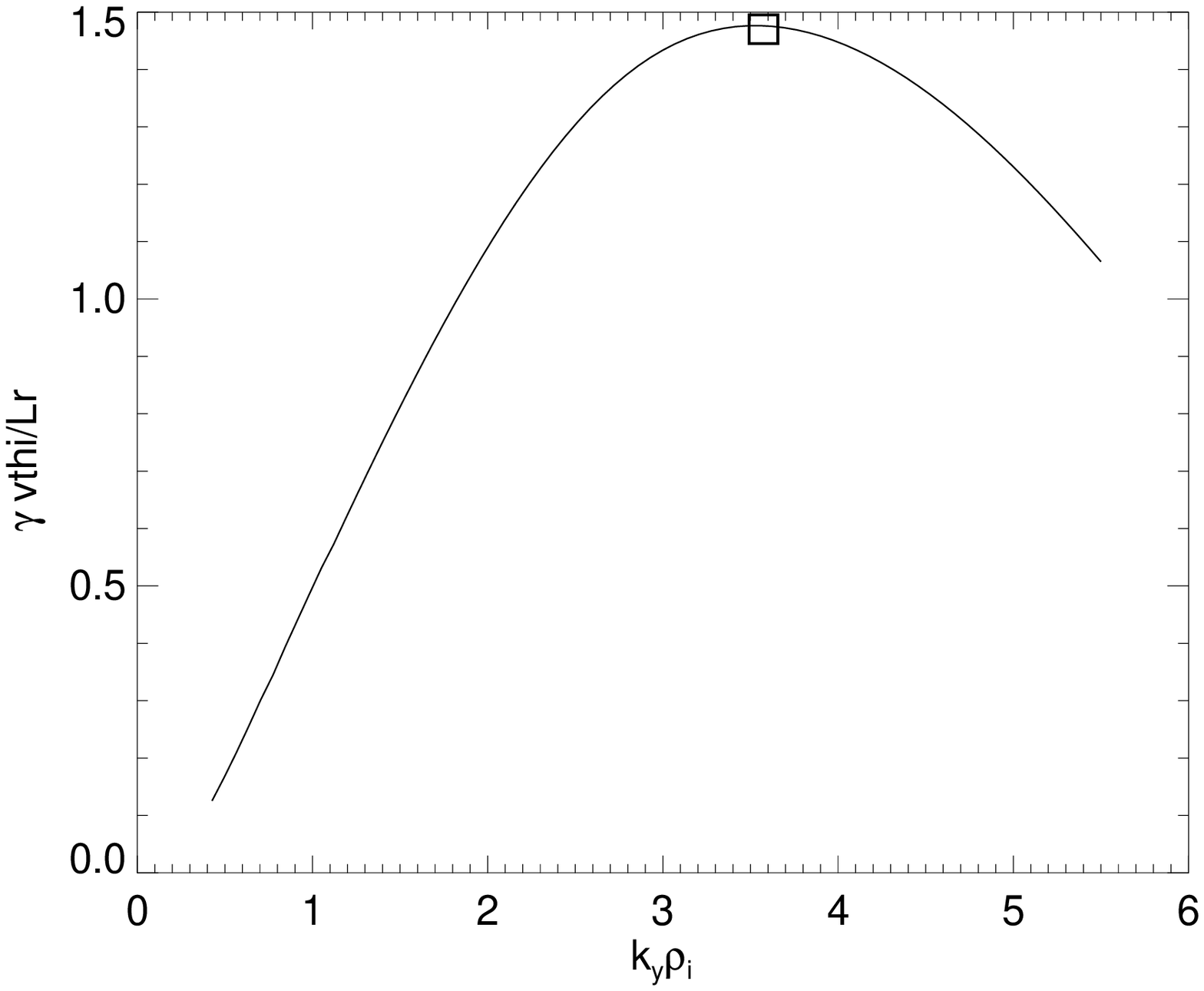}
		 	}
		}
		\hspace{0.001\columnwidth}
		\subfigure[]{
		    \psfrag {0.2} [cc][cc][1][0] {0.2 }
		    \psfrag {0.4} [cc][cc][1][0] {0.4 }
		    \psfrag {0.6} [cc][cc][1][0] {0.6 }
		    \psfrag {0.8} [cc][cc][1][0] {0.8 }
		    \psfrag {1.0} [cc][cc][1][0] {1.0 }
		    \psfrag {k} [cc][cc][1][0]
		    {\raisebox{-2.5ex}{\hspace{1em}\ky$\rho_i$}}
		    \psfrag {y} [cc][cc][1][0] {}
	    	\psfrag {r} [cc][cc][1][0] {}
	    	\psfrag {i} [cc][cc][1][0] {}
	    	\psfrag {0.0} [cc][cc][1][0] {0.0 }
	    	\psfrag {g} [cc][cc][1][0] 
{\raisebox{2ex}{\hspace{2em}$\gamma \brac{\vthi/\lref}$}}
		    \psfrag {v} [cc][cc][1][0] {}
		    \psfrag {th} [cc][cc][1][0] {}
	    	\psfrag {/L} [cc][cc][1][0] {}
	    	\psfrag {ref} [cc][cc][1][0] {}
		    \label{fig:Expt:CycSpectra:Gamma}
			\raisebox{2.5ex}{
				\includegraphics[width=0.45\columnwidth]
		 		{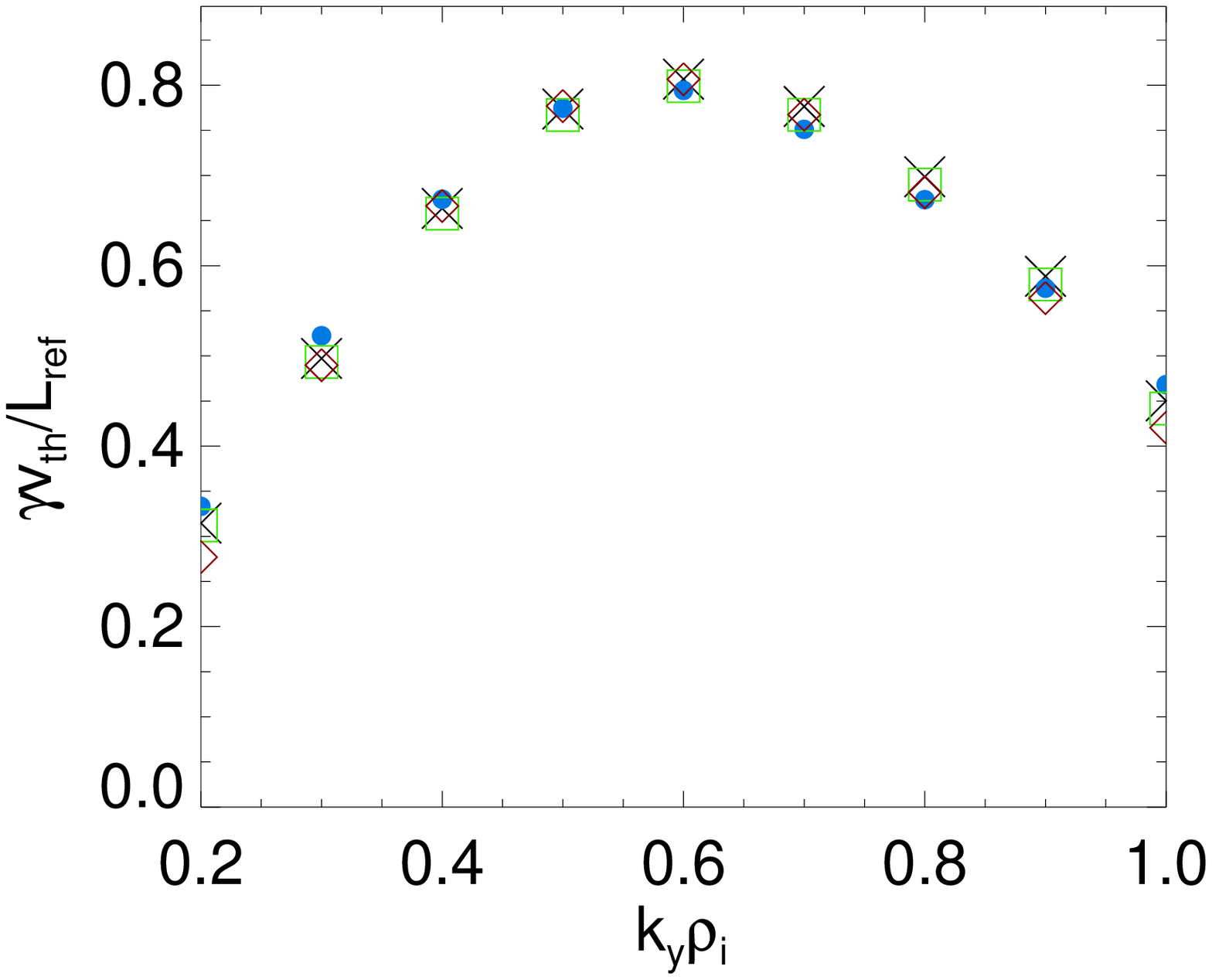}
		 	}
		}
			\subfigure[]{
				\psfrag {0.0} [cc][cc][0.9][0] {0.0}
			    \psfrag {0.5} [cc][cc][0.9][0] {0.5}
			    \psfrag {1.0} [cc][cc][0.9][0] {1.0}
			    \psfrag {1.5} [cc][cc][0.9][0] {1.5}
			    \psfrag {2.0} [cc][cc][0.9][0] {2.0}
			    \psfrag {R (m)} [cc][cc][1][0] 
			    	{\raisebox{-2.5ex}{$R$~(m)}}
			    \psfrag {-1.5} [cc][cc][0.9][0] {-1.5}
			    \psfrag {-1.0} [cc][cc][0.9][0] {-1.0}
			    \psfrag {-0.5} [cc][cc][0.9][0] {-0.5}
			    \psfrag {Z (m)} [cc][cc][1][0] {$Z$~(m)}
			    \psfrag {R=0.86} [cc][bc][0.9][0] 
			    	{ $R_0=0.86$}
			    \psfrag {r=0.69} [cc][bc][0.9][30] {$r=0.69$}
			    \label{fig:Expt:SurfCompare}
				\includegraphics[width=0.35\columnwidth]
					{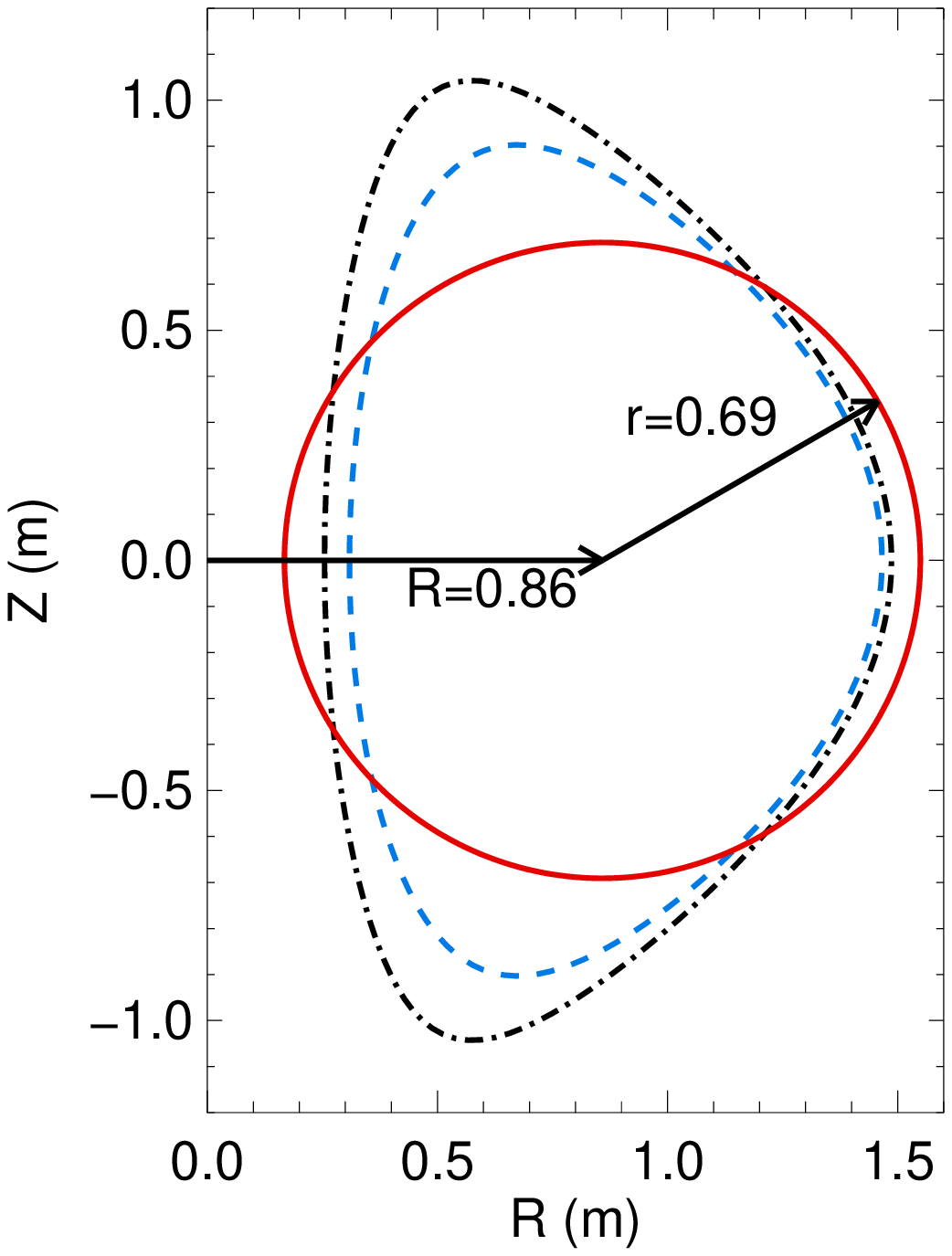}		
			}
        }
    \caption[Growth rate spectrum for MAST flux surface.]
     {
(a) $\gmtm$ spectrum for the MAST flux surface at $\psi_N=0.94$, highlighting the peak wavenumber (\opensquare{}) corresponding to the $\phi$ eigenfunction of \fig{fig:Intro:CoreEdgePhiComp:Edge}.
(b) $\gmtm$ spectra for the circular equilibrium fit [{\color{red} \full} in (c)] with: the standard full physics model excluding $B_\|$ ({\color{dkred} $\opendiamond$}); with adiabatic ions ({\color{black} $\times$}); including $B_\|$ ({\color{green} \opensquare{}}) and neglecting $\phi$ ({\color{blue} \fullcircle{}}).
(c) Circular fit ({\color{red} \full}) to the $\psi_N=0.94$ flux surface ({\color{blue} \dashed}) along with the last closed flux surface (\chain).
}
    \label{fig:Expt:Spectra}
\end{figure}

We base our studies on a reference local equilibrium from the plateau region at the top of a MAST H-mode pedestal, which is unstable to MTMs \footnote{A full account of the equilibrium reconstruction from MAST data is given in \Ref{Dickinson2011}.}. 
The reference flux surface is $\psi_N=0.94$ at the midpoint during the ELM cycle, with the equilibrium parameters given in \tab{tab:ExptSurfPar}.
The growth rate spectrum peaks at $\ky\rhoi{}\sim3.5$ \footnotemark{}, and is shown in \fig{fig:Expt:Spectra:Gamma}.
\footnotetext{MTMs with peak growth rate at $\ky\rhoi\gg1$ have also been found to dominate close to the core of NSTX plasmas \cite{Smith2011}.}
The minimal equilibrium conditions necessary to drive MTMs unstable are sought by progressively simplifying the equilibrium assumptions.

Sensitivity to flux surface shaping is investigated by fitting the reference MAST equilibrium using the simple \sa{} shifted concentric circle model \cite{Connor1978}. 
This model allows easy independent control of the main equilibrium parameters: safety factor $q$; magnetic shear $\shat=rq^\prime/q$ (where $^{\prime}=d/dr$); inverse aspect ratio $\epsilon=r/R$ (which sets the trapped fraction); normalised pressure gradient $\alpha=$$Rq^2\beta/L_p$; normalised inverse temperature and density gradient scale lengths $\lref/\lte$, $\lref/\lne$; and magnetic drift strength parameter $\epsilon_l=2\lref/R$.
The circle is a crude fit to the edge of MAST, as illustrated in \fig{fig:Expt:SurfCompare} which compares this fit with the experimental flux surface.
The $\gmtm$ spectrum for the circular fit is shown in \fig{fig:Expt:CycSpectra:Gamma}.
There is a significant shift in the \ky\rhoi{} at which $\gmtm$ peaks relative to the shaped surface case, but the magnitudes of $\gmtm$ (and $\omega$) are within a factor 2.
This is consistent with previous studies showing that shaping is not essential for MTMs \cite{Applegate2007}.

%
\begin{table}
\caption{\label{tab:ExptSurfPar}Equilibrium parameters characteristic of MAST shot \#24763 at the mid-point in time between two ELMs for $\psi_N=0.94$.
$^{\dagger}$ NB $\nue$ is normalised to $\vthi/\lref$.}
\begin{center}
\begin{tabular}{@{}lllllllll}
	\br
	$q$ &$\shat$&$\epsilon$&$\epsilon_l$&$\lref/\lte$&
	$\lref/\lne$&$\beta_e$
	&$\alpha$&$\nue$ $^{\dagger}$\\
	\mr
	4.66&7.67&0.805&1.435&5.88&0.36&0.015&-5.66&1.98\\
	\br
\end{tabular}
\end{center}
\end{table}

%

%

%

In this \sa{} model equilibrium, \fig{fig:Expt:CycSpectra:Gamma} shows that calculations with fully kinetic and purely Boltzmann ion responses yield very similar $\gmtm$ spectra.
Previous simulations of MTMs in the core also found that $\gmtm$ is insensitive to including fully kinetic ions as the ion response is close to Boltzmann \cite{Applegate2007,Told2008,Guttenfelder2012}.
In early treatment of collisionless and semi-collisional MTMs the kinetic ion response was neglected  \cite{Drake1977}, but was found to be strongly stabilising when \rhoi$>d$ \cite{Cowley1986}, where $d$ is the width of the current layer associated with the mode.
Inspection of the \apar{} eigenfunction, for the dominant MTM in this \sa{} equilibrium, provides an estimate of the current layer width, $d \sim 0.4\rhoi$.
If the current layer width were as narrow as $\delta_0$, estimated from \eqn{eq:Intro:SemiCollSkinDepth} as $\delta_0/\rhoi\sim\order{10^{-2}}$, the kinetic ion response would be expected to be stabilising in the model of \Ref{Cowley1986}, but this stabilising effect was not observed in our simulations with kinetic ions. 

\Fig{fig:Expt:CycSpectra:Gamma} also shows that the growth rate is insensitive to including compressional magnetic perturbations, $B_\|$, and only weakly sensitive to including the electrostatic potential, $\phi$, which has a modest impact on the frequency spectrum (not shown).
Subsequent simulations in this paper will use the \sa{} equilibrium model, retain $\phi$, and neglect the kinetic ion response and $B_\|$.

The studies of \sect{sec:ModeChar} are based on scans around the reference equilibium, during which MTMs can become subdominant to other instabilities.
GS2 is an initial value code, and subdominant MTMs are tracked in this up-down symmetric equilibrium, by filtering to keep only the component of the nonadiabatic perturbed distribution function with odd parity in the parallel direction.
These are tearing parity modes (i.e. modes where $\phi$ is odd and \apar{} is even about $\theta=0$).
Finally both the semi-collisional width, $\delta_0$, and the collisionless width, $\delta_n=\rho_e\sqrt{2/\beta_e}$, are resolved by using a domain that is sufficiently extended in $\theta$, ($-11\pi< \theta < 11\pi$).

\section{Linear mode analysis}
\label{sec:ModeChar}
There have been several linear gyrokinetic studies of how MTM stability depends on equilibrium parameters \cite{Applegate2007,Told2008,Guttenfelder2012}.
Here we explore a new region of parameter space, by moving to extremely low aspect ratio and high magnetic shear, which characterises the MAST edge.

\subsection{Temperature and density dependence}
A finite electron temperature gradient is essential for both MTM drive mechanisms described in \sect{sec:Intro}, with the onset of instability arising above a threshold gradient.
The MTM's real frequency, $\omega$, is predicted to vary approximately linearly with the electron diamagnetic frequency, $\wde$, with a precise relationship that depends on the driving mechanism.

\begin{figure}[htb]
    \centering
    	\subfigure[]{
    	    \psfrag {0} [cc][cc][1][0] {0}
		    \psfrag {2} [cc][cc][1][0] {2}
		    \psfrag {4} [cc][cc][1][0] {4}
		    \psfrag {6} [cc][cc][1][0] {6}
		    \psfrag {8} [cc][cc][1][0] {8}
		    \psfrag {10} [cc][cc][1][0] {10}
		    \psfrag {12} [cc][cc][1][0] {12}
		    \psfrag {L} [cc][cb][1][0] 
		    	{\raisebox{-2.5ex}{\hspace{3em}\lref{}/\lte{}}}
		    \psfrag {ref} [cc][cc][1][0] {}
		    \psfrag {/L} [cc][cc][1][0] {}
		    \psfrag {Te} [cc][cc][1][0] {}
		    \psfrag {0.2} [cc][cc][1][0] {0.2}
		    \psfrag {0.4} [cc][cc][1][0] {0.4}
		    \psfrag {0.6} [cc][cc][1][0] {0.6}
		    \psfrag {0.8} [cc][cc][1][0] {0.8}
		    \psfrag {1.0} [cc][cc][1][0] {1.0}
		    \psfrag {k} [cc][cc][1][0] 		    
		    	{\hspace{1em}\ky\rhoi{}}
		    \psfrag {y} [cc][cc][1][0] {}
		    \psfrag {r} [cc][cc][1][0] {}
		    \psfrag {i} [cc][cc][1][0] {}
		    \psfrag { } [cc][cc][1][0] { }
		    \psfrag {-1.50} [cc][cc][1][0] {-1.5}
		    \psfrag {-1.00} [cc][cc][1][0] 
		    	{}
		    \psfrag {-0.50} [cc][cc][1][0] 
		    	{}
		    \psfrag {0.00} [cc][cc][1][0] {\hspace{1em}0.0}
		    \psfrag {0.50} [cc][cc][1][0] 
		    	{}
		    \psfrag {1.00} [cc][cc][1][0] 
		    	{}
		    \psfrag {1.50} [cc][cc][1][0] {\hspace{1em}1.5}
		    \psfrag {g} [cc][cc][1][0] 
		    	{\hspace{4em}$\gamma \brac{\vthi/\lref}$}
		    \psfrag {v} [cc][cc][1][0] {}
		    \psfrag {th} [cc][cc][1][0] {}
		    \psfrag {} [cc][cc][1][0] {}
			\label{fig:Lin:Gamma:Tprim}
			\includegraphics[width=0.47\columnwidth]
		 	{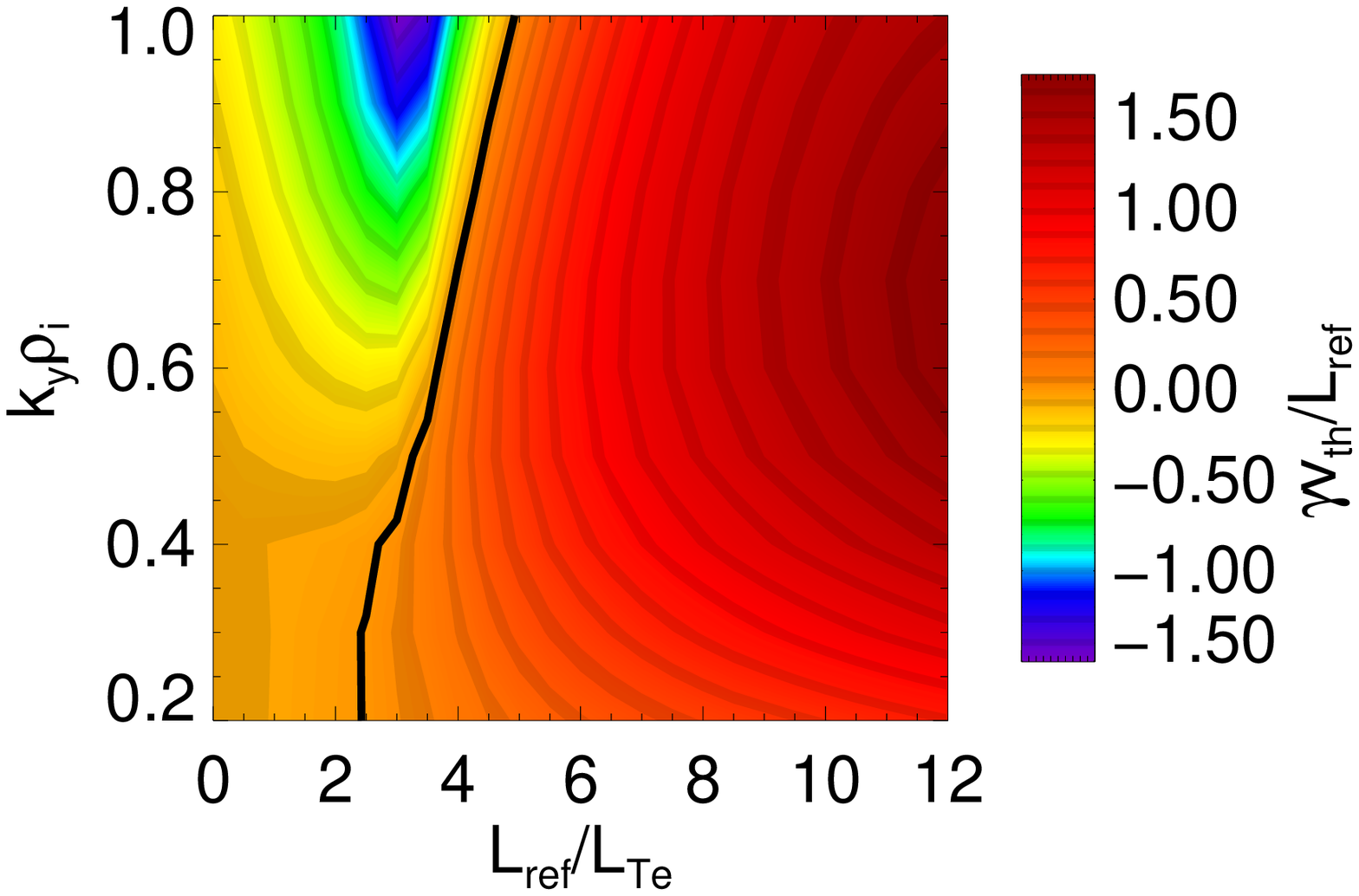}
		}
        \subfigure[]{
            \psfrag {-2} [cc][cc][1][0] {-2}
		    \psfrag {0} [cc][cc][1][0] {0}
		    \psfrag {2} [cc][cc][1][0] {2}
		    \psfrag {4} [cc][cc][1][0] {4}
		    \psfrag {6} [cc][cc][1][0] {6}
		    \psfrag {8} [cc][cc][1][0] {8}
		    \psfrag {10} [cc][cc][1][0] {10}
		    \psfrag {12} [cc][cc][1][0] {12}
		    \psfrag {L} [cc][cb][1][0] 
	    	{\raisebox{-2.5ex}{\hspace{3em}\lref{}/\lne{}}}
		    \psfrag {ref} [cc][cc][1][0] {}
		    \psfrag {/L} [cc][cc][1][0] {}
		    \psfrag {ne} [cc][cc][1][0] {}
		    \psfrag {0.2} [cc][cc][1][0] {0.2}
		    \psfrag {0.4} [cc][cc][1][0] {0.4}
		    \psfrag {0.6} [cc][cc][1][0] {0.6}
		    \psfrag {0.8} [cc][cc][1][0] {0.8}
		    \psfrag {1.0} [cc][cc][1][0] {1.0}
		    \psfrag {k} [cc][cc][1][0] 
		    	{\hspace{1em}\ky\rhoi{}}
		    \psfrag {y} [cc][cc][1][0] {}
		    \psfrag {r} [cc][cc][1][0] {}
		    \psfrag {i} [cc][cc][1][0] {}
		    \psfrag { } [cc][cc][1][0] { }
		    \psfrag {-0.50} [cc][cc][1][0] {-0.5}
		    \psfrag {0.00} [cc][cc][1][0] {\hspace{1em}0.0}
		    \psfrag {0.50} [cc][cc][1][0] {\hspace{1em}0.5}
		    \psfrag {1.00} [cc][cc][1][0] {\hspace{1em}1.0}
		    \psfrag {g} [cc][cc][1][0] 
		    	{\hspace{4em}$\gamma \brac{\vthi/\lref}$}
		    \psfrag {v} [cc][cc][1][0] {}
		    \psfrag {th} [cc][cc][1][0] {}
			\label{fig:Lin:Gamma:Fprim}
			\includegraphics[width=0.47\columnwidth]
		 	{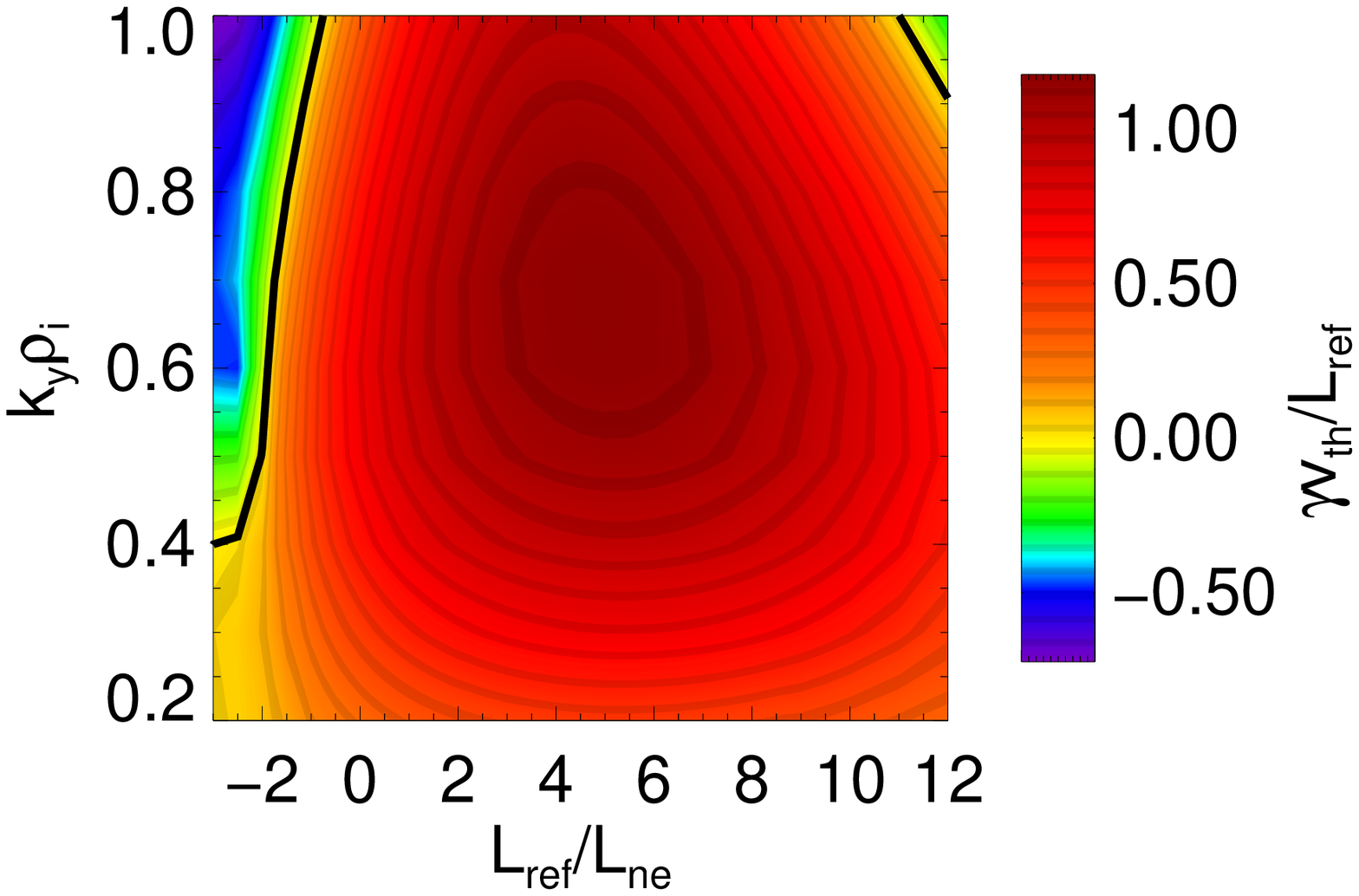}
		}
    \caption[Text]{
    The growth rate as a function of  $\ky\rhoi$ for varying (a) temperature and (b) density gradient length scales. 
    The location of marginality, $\gmtm=0$, is given (\full).
    }
    \label{fig:Lin:Spectra:Gamma}
\end{figure}

Scans have been performed by varying the normalised gradient length scales, $\lref/\lte$ and $\lref/\lne$, independently, at fixed values of all other parameters \footnotemark{}.
\footnotetext{The normalised pressure gradient, $\alpha$, was held constant in these scans.}
The resulting growth rate spectrum in \fig{fig:Lin:Gamma:Tprim} shows a finite threshold temperature gradient that increases with \ky{}, and that $\gmtm$ increases monotonically above this threshold.
\Fig{fig:Lin:Gamma:Fprim} shows that MTMs are unstable at $\lref/\lne=0$, and that $\gmtm$ is maximised at finite $\lref/\lne$ similar to previous findings \cite{Applegate2007,Guttenfelder2012}.
In both scans $\omega$ is found to be reasonably well described by $\wde\brac{a+b\eta_e}$ where $\eta_e=\lne/\lte$, in qualitative agreement with analytic predictions.

%
%

\subsection{Beta}
$\gmtm$ is insensitive to $\phi$, and MTMs are driven by the magnetic perturbation, \apar{}.
Therefore $\gmtm$ will be strongly affected by $\beta$, which controls the strength of magnetic perturbations through Amp\`eres law.
Electromagnetic instabilities, like kinetic ballooning modes and MTMs, are typically unstable above a threshold $\beta$, with growth rates that then increase strongly with $\beta$ \cite{Snyder2001}.
This may explain discrepancies between different tokamaks in the observed confinement scaling with $\beta$ \cite{Petty2008}: increases in $\beta$ that cross the threshold will increase transport whilst increases that remain below the threshold will have less impact (and may stabilise other instabilities \cite{Tang1985,Belli2010}).

\begin{figure}[htb]
    \centering
    \resizebox{1.0\columnwidth}{!}{
    	\subfigure[]{
    	    \psfrag {0.01} [cc][cc][0.9][0] {0.01}
		    \psfrag {0.02} [cc][cc][0.9][0] {0.02}
		    \psfrag {0.03} [cc][cc][0.9][0] {0.03}
		    \psfrag {b} [cc][cc][1][0] 
		    	{\raisebox{-2.5ex}{$\beta_e$}}
		    \psfrag {e} [cc][cc][1][0] {}
		    \psfrag {0.0} [cc][cc][0.9][0] {0.0}
		    \psfrag {0.2} [cc][cc][0.9][0] {0.2}
		    \psfrag {0.4} [cc][cc][0.9][0] {0.4}
		    \psfrag {0.6} [cc][cc][0.9][0] {0.6}
		    \psfrag {0.8} [cc][cc][0.9][0] {0.8}
		    \psfrag {1.0} [cc][cc][0.9][0] {1.0}
		    \psfrag {g} [cc][cc][1][0] 
		    	{\hspace{4em}$\gamma \brac{\vthi/\lref}$}
		    \psfrag {v} [cc][cc][1][0] {}
		    \psfrag {th} [cc][cc][1][0] {}
		    \psfrag {} [cc][cc][1][0] {}
		    \psfrag {/L} [cc][cc][1][0] {}
		    \psfrag {ref} [cc][cc][1][0] {}
			\label{fig:Lin:Gamma:Beta}
			\includegraphics[width=0.4\columnwidth]
		 	{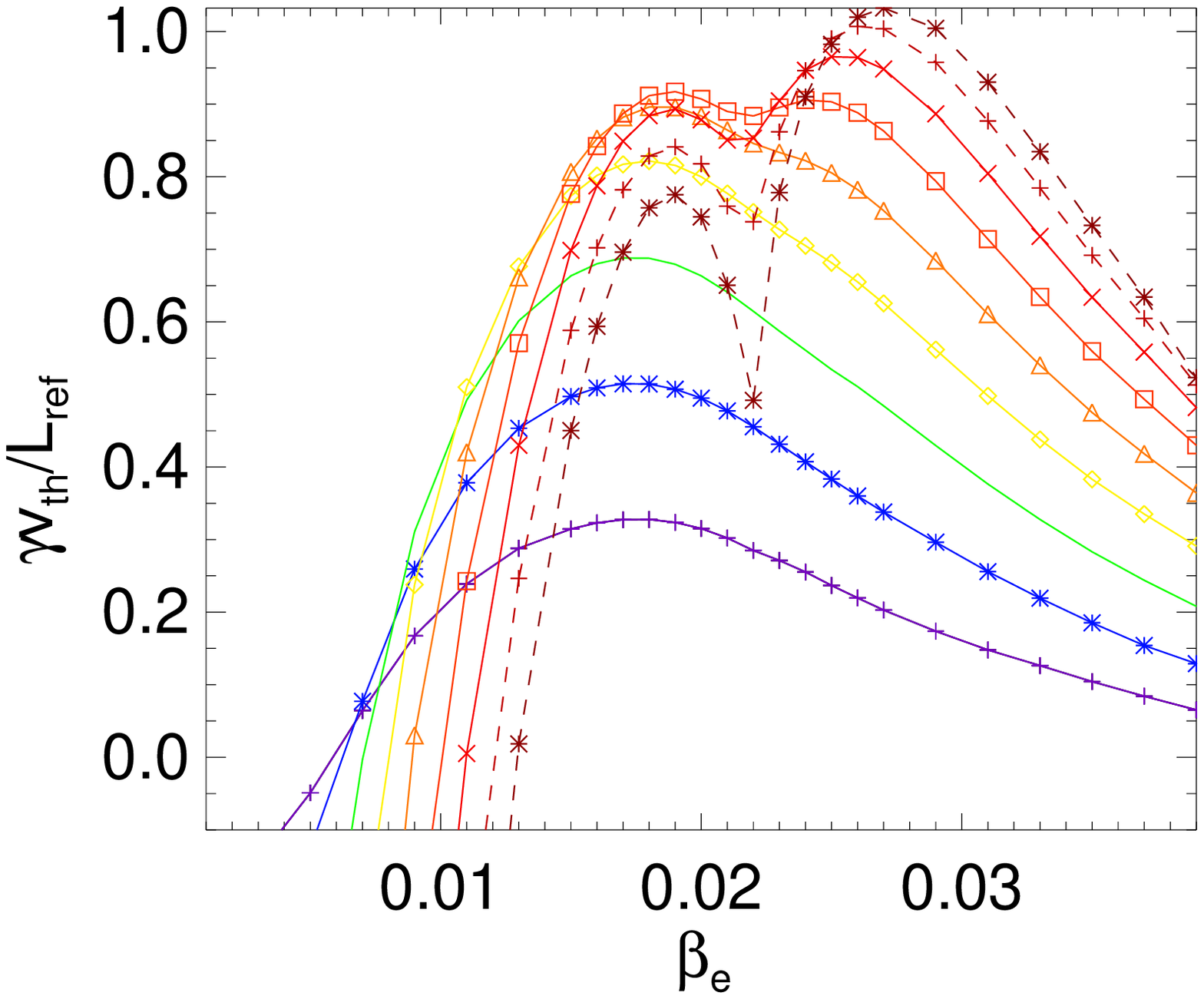}
		}
        \hspace{-1em}
    	\subfigure[]{
    	    \psfrag {1x10} [cc][cc][0.8][0] {1$\times$10}
		    \psfrag {-2} [cc][cc][0.6][0] {\hspace{0.3em}-2}
		    \psfrag {-1} [cc][cc][0.6][0] {\hspace{0.3em}-1}
		    \psfrag {0} [cc][cc][0.6][0] {\hspace{0.3em}0}
		    \psfrag {1} [cc][cc][0.6][0] {\hspace{0.3em}1}
		    \psfrag {2} [cc][cc][0.6][0] {\hspace{0.3em}2}
		    \psfrag {3} [cc][cc][0.6][0] {\hspace{0.3em}3}
		    \psfrag {n} [cc][cc][1][0] 
		    	{\raisebox{-2.5ex}{\hspace{5em}$\nue \brac{\vthi/\lref}$}}
		    \psfrag {ei} [cc][cc][1][0] {}
		    \psfrag {v} [cc][cc][1][0] {}
		    \psfrag {th} [cc][cc][1][0] {}
		    \psfrag {/L} [cc][cc][1][0] {}
		    \psfrag {ref} [cc][cc][1][0] {}
		    \psfrag {} [cc][cc][1][0] {}
		    \psfrag {0.0} [cc][cc][0.9][0] {0.0}
		    \psfrag {0.2} [cc][cc][0.9][0] {0.2}
		    \psfrag {0.4} [cc][cc][0.9][0] {0.4}
		    \psfrag {0.6} [cc][cc][0.9][0] {0.6}
		    \psfrag {0.8} [cc][cc][0.9][0] {0.8}
		    \psfrag {g} [cc][cc][1][0] 
		    	{\hspace{4em}$\gamma \brac{\vthi/\lref}$}
			\label{fig:Lin:Gamma:Coll}
			\includegraphics[width=0.42\columnwidth]
		 	{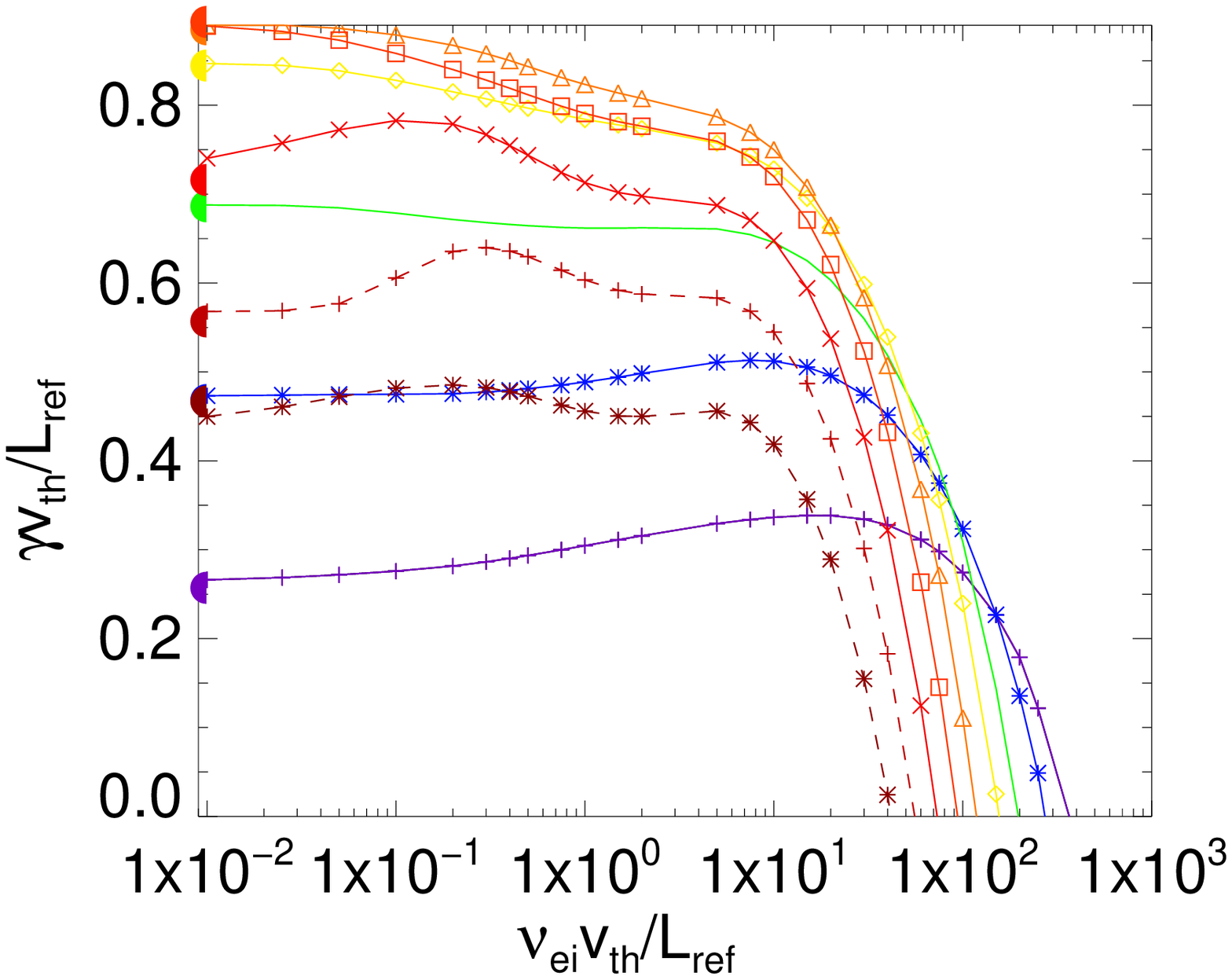}
		}
		\hspace{-2.2em}
        \subfigure{
            \psfrag {k} [cc][cc][1][0] 	
            	{\hspace{1em}\ky\rhoi{}}
		    \psfrag {y} [cc][cc][1][0] {}
		    \psfrag {r} [cc][cc][1][0] {}
		    \psfrag {i} [cc][cc][1][0] {}
		    \psfrag {=0.2} [cc][cc][1][0] 
		    	{\hspace{1.8em}=0.2}
		    \psfrag {=0.3} [cc][cc][1][0] 
		    	{\hspace{1.8em}=0.3}
		    \psfrag {=0.4} [cc][cc][1][0] 
		    	{\hspace{1.8em}=0.4}
		    \psfrag {=0.5} [cc][cc][1][0] 
		    	{\hspace{1.8em}=0.5}
		    \psfrag {=0.6} [cc][cc][1][0] 
		    	{\hspace{1.8em}=0.6}
		    \psfrag {=0.7} [cc][cc][1][0] 
		    	{\hspace{1.8em}=0.7}
		    \psfrag {=0.8} [cc][cc][1][0] 
		    	{\hspace{1.8em}=0.8}
		    \psfrag {=0.9} [cc][cc][1][0] 
		    	{\hspace{1.8em}=0.9}
		    \psfrag {=1.0} [cc][cc][1][0] 
		    	{\hspace{1.8em}=1.0}
			\label{fig:Lin:KyKey}
			\raisebox{5ex}{
				\includegraphics[width=0.12\columnwidth]
		 		{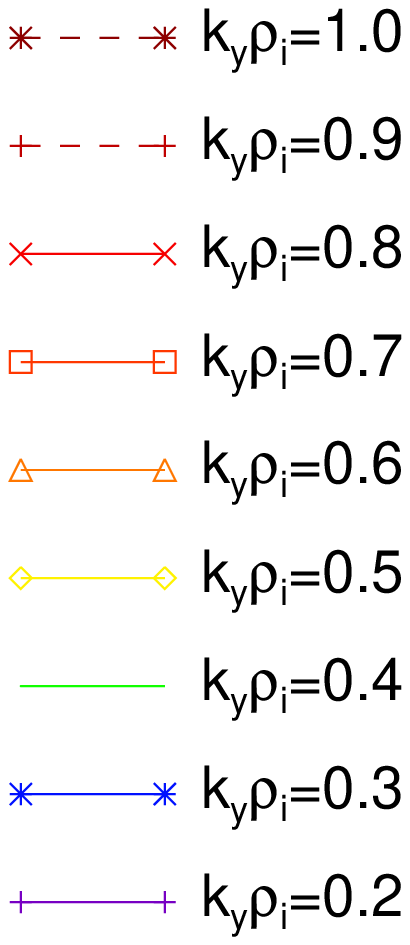}
		 	}
		 }
	}
    \caption[Text]{
    $\gmtm$ as a function of (a) $\beta_e$ and (b) $\nue$, for modes over a range of \ky\rhoi{} values.
    During the $\beta_e$ scan the normalised pressure gradient, $\alpha$, is varied consistently. 
    The results of simulations with $\nue=0$ are shown in (b) by the filled semi-circles on the y-axis, indicating a substantial growth rate even in the absence of collisions. 
    The key to the $\ky\rhoi$ values also applies to \figs{fig:Lin:Spectra:Eps},~\ref{fig:Lin:Spectra:Q} and~\ref{fig:Lin:Gamma:Epsl}.
    }
    \label{fig:Lin:Spectra:BetaAndColl}
\end{figure}

The growth rates are shown for a range of \ky{} values in \fig{fig:Lin:Gamma:Beta} for a scan in $\beta_e$ where $\alpha$ is scaled consistently.
There is a clear stability threshold in $\beta_e$, which increases approximately linearly with \ky{}, above which $\gmtm$ increases rapidly with $\beta_e$.
For sufficiently high $\beta_e$, further increases in $\beta_e$ become stabilising, as was also seen in \Ref{Applegate2007}.
This stabilisation at high $\beta_e$ is stronger when $\alpha$ is scaled consistently than if $\alpha$ is fixed, which is consistent with magnetic drifts becoming more favourable at higher $\alpha$ \cite{Roach1995}. 
The local minimum in $\gmtm$ at $\beta_e\sim0.022$ is only seen in the scan with $\alpha$ varying consistently, and not with $\alpha$ fixed.

%
%

\subsection{Collision frequency}
It has already been pointed out that collisions play an essential role in the existing analytic drive mechanisms for MTMs. 
Linear gyrokinetic simulations have generally reported growth rates that peak for $\nub\sim\order{1}$, as may be expected from a mode driven by a combination of slab and trapped particle drives.
Recent simulations have shown $\gmtm$ dropping by only a factor of 2, as $\nue$ falls by over two orders of magnitude from its value at the peak \cite{Doerk2012}, which suggests that as the collision based drive is removed, a further substantial drive mechanism remains.

\Fig{fig:Lin:Gamma:Coll} shows $\gmtm$ as a function of $\nue$ for a range of \ky\rhoi{} values.
Increasing the collision frequency well above $\nub\sim\order{1}$ is stabilising.
It is more striking that $\gmtm$ for the dominant mode, and at several other values of \ky{}, does not peak at finite $\nue$, but remains constant or even slowly increases as $\nue$ decreases all the way to zero, which is in stark contrast to the ``usual'' core behaviour where $\gmtm$ peaks at $\nub \sim \order{1}$ (e.g. at mid-radius in MAST \cite{Applegate2007}) \footnotemark{}.
\footnotetext{$\gmtm$ for the lowest \ky{} mode peaks at finite $\nue$, and drops only slowly with decreasing $\nue$, closely resembling the dependence presented in \Ref{Doerk2012}.} 
In \ref{sec:NumDiss} it is demonstrated that this collision frequency dependence is robustly reproduced using grids with higher resolutions in velocity space.
The trapped particle drive mechanism of \Ref{Catto1981a} requires collisions and must vanish at $\nue \equiv 0$: it therefore cannot be responsible for the instability seen here.
A {\bf collisionless} mechanism is required, which cannot rely on the time-dependent thermal force.

A collision frequency scan for the fully shaped MAST edge equilibrium also finds that $\gmtm$ peaks at $\nue \sim 0$, as for the edge \sa{} model equilibrium and in contrast to the $\nue$ dependence at mid-radius.
Could the different dependences of $\gmtm$ on collision frequency be explained by substantial differences between the core and edge values of inverse aspect ratio, $\epsilon=r/R$, and magnetic shear, $\shat$?

\subsection{Aspect ratio (trapped particles)}
\begin{figure}[htb]
    \centering
    \resizebox{1.0\columnwidth}{!}{
    	\subfigure[]{
    	    \psfrag {0.0} [cc][cc][0.9][0] {0.0}
		    \psfrag {0.2} [cc][cc][0.9][0] {0.2}
		    \psfrag {0.4} [cc][cc][0.9][0] {0.4}
		    \psfrag {0.6} [cc][cc][0.9][0] {0.6}
		    \psfrag {0.8} [cc][cc][0.9][0] {0.8}
		    \psfrag {e} [cc][cc][1][0] 	
		    	{\raisebox{-2.5ex}{$\epsilon$}}
		    \psfrag {-0.2} [cc][cc][0.9][0] {-0.2}
		    \psfrag {g} [cc][cc][1][0] 
		    	{\hspace{4em}$\gamma \brac{\vthi/\lref}$}
		    \psfrag {v} [cc][cc][1][0] {}
		    \psfrag {th} [cc][cc][1][0] {}
		    \psfrag {} [cc][cc][1][0] {}
		    \psfrag {/L} [cc][cc][1][0] {}
		    \psfrag {ref} [cc][cc][1][0] {}
			\label{fig:Lin:Gamma:Eps}
			\includegraphics[width=0.4\columnwidth]
		 	{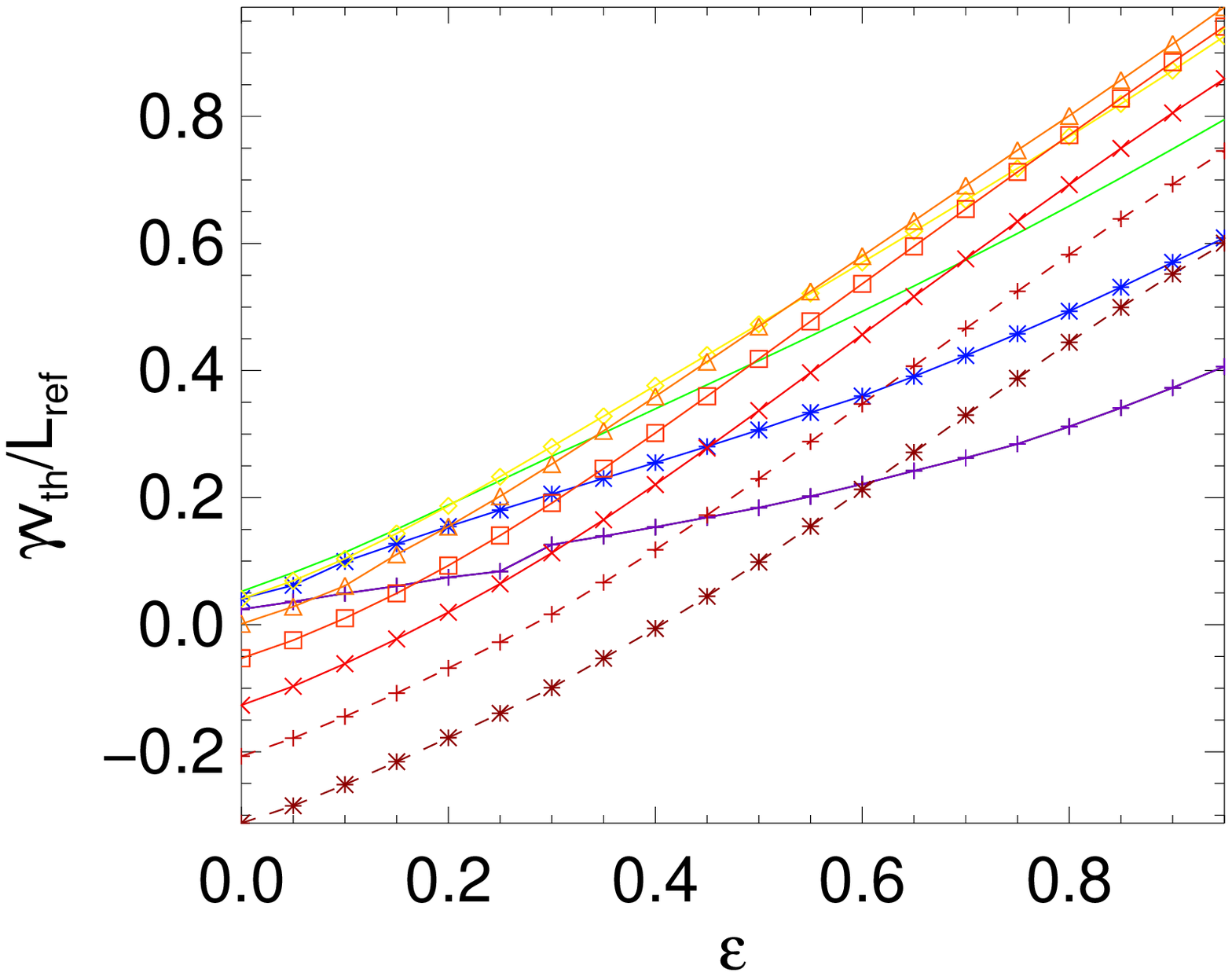}
		}
        \hspace{0.001\columnwidth}
		\subfigure[]{
            \psfrag {1x10} [cc][cc][0.8][0] {1$\times$10}
		    \psfrag {-2} [cc][cc][0.6][0] {\hspace{0.3em}-2}
		    \psfrag {-1} [cc][cc][0.6][0] {\hspace{0.3em}-1}
		    \psfrag {0} [cc][cc][0.6][0] {\hspace{0.3em}0}
		    \psfrag {1} [cc][cc][0.6][0] {\hspace{0.3em}1}
		    \psfrag {2} [cc][cc][0.6][0] {\hspace{0.3em}2}
			\psfrag {n} [cc][cc][1][0] 
		    	{\raisebox{-2.5ex}{\hspace{5em}$\nue \brac{\vthi/\lref}$}}
		    \psfrag {ei} [cc][cc][1][0] {}
		    \psfrag {v} [cc][cc][1][0] {}
		    \psfrag {th} [cc][cc][1][0] {}
		    \psfrag {/L} [cc][cc][1][0] {}
		    \psfrag {ref} [cc][cc][1][0] {}
		    \psfrag {} [cc][cc][1][0] {}
    	    \psfrag {0.0} [cc][cc][0.9][0] {0.0}
		    \psfrag {0.2} [cc][cc][0.9][0] {0.2}
		    \psfrag {0.4} [cc][cc][0.9][0] {0.4}
		    \psfrag {0.6} [cc][cc][0.9][0] {0.6}
		    \psfrag {0.8} [cc][cc][0.9][0] {0.8}
		    \psfrag {e} [cc][cc][1][0] {$\epsilon$}
		    \psfrag { } [cc][cc][1][0] { }
		    \psfrag {-0.40} [cc][cc][1][0] {-0.40}
		    \psfrag {-0.20} [cc][cc][1][0] {}
		    \psfrag {-0.00} [cc][cc][1][0] 
		    	{\phantom{-}0.00}
		    \psfrag {0.20} [cc][cc][1][0] 
		    	{}
		    \psfrag {0.40} [cc][cc][1][0] 
		    	{\hspace{1em}0.40}
		    \psfrag {0.60} [cc][cc][1][0] 
		    	{}
		    \psfrag {0.80} [cc][cc][1][0] 
		    	{\hspace{1em}0.80}
		    \psfrag {1.00} [cc][cc][1][0] 	
		    	{}
		    \psfrag {g} [cc][cb][1][0] 
		    	{\hspace{3em}$\gamma \brac{\vthi/\lref}$}
		    \label{fig:Lin:Spectra:EpsColl}
    	\includegraphics[width=0.48\columnwidth]
		 	{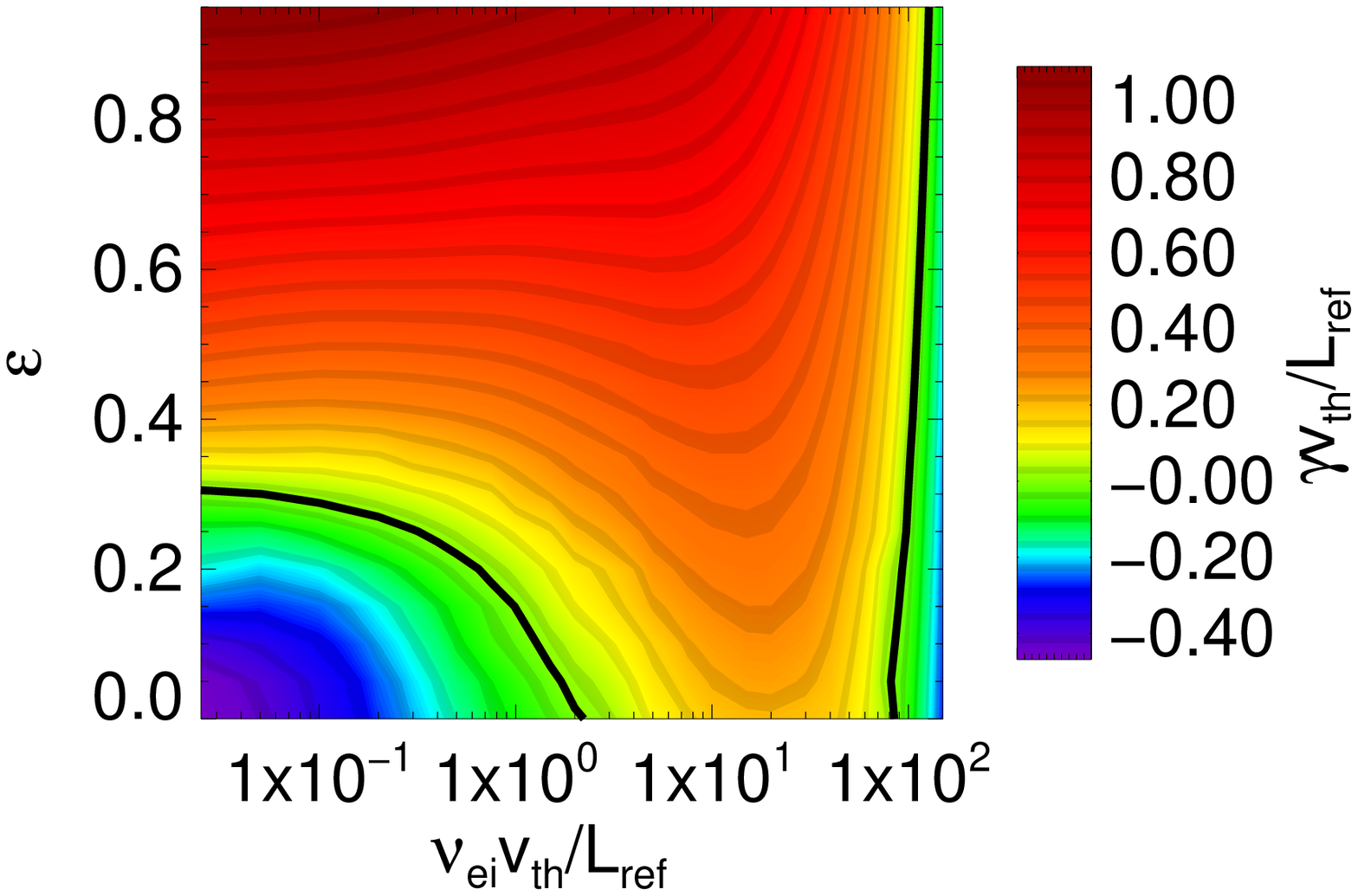}
		}
	}
    \caption[Text]{
    (a) $\gmtm$ as a function of $\epsilon$ for each of the \ky\rhoi{} values in the key of \fig{fig:Lin:Spectra:BetaAndColl} and with $\nue=1.98$.
    (b) $\gmtm$ for the dominant $\ky\rhoi=0.6$ mode as a function of $\epsilon$ and $\nue$, with \full{} indicating marginal stability.
    The frequency, $\omega$, remains between -1.5 and -2.8 throughout these scans.
    }
    \label{fig:Lin:Spectra:Eps}
\end{figure}

Varying only the inverse aspect ratio, $\epsilon=r/R$, in the \sa{} model, corresponds to changing the trapped particle fraction whilst holding all other parameters fixed.
The results from this scan, illustrated in \fig{fig:Lin:Gamma:Eps}, reveal a strong dependence of $\gmtm$ on $\epsilon$, and suggest that trapped particles are important to the linear drive for the reference value of $\nue$.
The decline in $\gmtm$ with decreasing $\epsilon$ is nearly uniform for $\ky\rhoi\geq0.5$, but rather weaker at lower \ky{}.
This suggests that at low \ky{} the trapped particle drive may be complemented by another mechanism at the nominal $\nue$, which would also be consistent with \fig{fig:Lin:Gamma:Coll}.

\Ref{Applegate2007} found that trapped particles are destabilising for $\nub \ll 1$, but stabilising for $\nub\gtrsim0.5$ which is the reference collisionality regime here.
Furthermore, the trapped particle drive was shown to be most effective at low $\epsilon$ (unlike in \fig{fig:Lin:Gamma:Eps}). 
This is consistent with the trapped particle drive mechanism of \cite{Catto1981a}, where the trapped-passing boundary provides an instability drive but the trapped electrons are themselves stabilising as, in this theory, they cannot carry the current perturbation. 
The situation is different for the MTMs studied here.
\Fig{fig:Lin:Gamma:Eps} suggests that trapped particles provide a direct MTM drive, and \fig{fig:Lin:Gamma:Coll} shows that this survives without collisions.

%

%

\Fig{fig:Lin:Spectra:EpsColl} shows how $\gmtm$ for the dominant mode, at $\ky\rhoi=0.6$, depends on $\nue$ and $\epsilon$.
The dependence of $\gmtm$ on $\epsilon$ is strongest at low $\nue$.
At low $\epsilon$ the growth rate maximises at $\nue \sim \order{10}$, but at high $\epsilon$ the growth rate peaks at the minimum $\nue$.
Indeed for $\epsilon<0.3$ the MTMs become stable for sufficiently small $\nue$, and a strong peak in $\gmtm$ is seen for $\nub\sim10$, which is consistent with previous findings \cite{Applegate2007,Guttenfelder2012,Doerk2012}.  
The role of $\epsilon$ in enhancing $\gmtm$ should be most important near the pedestal region where the trapped particle fraction is maximised, especially in STs where $\epsilon$ approaches 1.

\subsection{Safety factor and magnetic shear}
%
\begin{figure}[htb]
    \centering
    \resizebox{1.0\columnwidth}{!}{
    	\subfigure[]{
			\psfrag {2} [cc][cc][0.9][0] {2}
		    \psfrag {4} [cc][cc][0.9][0] {4}
		    \psfrag {6} [cc][cc][0.9][0] {6}
		    \psfrag {8} [cc][cc][0.9][0] {8}
		    \psfrag {q} [cc][cc][1][0] 
		    	{\raisebox{-2.5ex}{q}}
		    \psfrag {-0.5} [cc][cc][0.9][0] {-0.5}
		    \psfrag {0.0} [cc][cc][0.9][0] {0.0}
		    \psfrag {0.5} [cc][cc][0.9][0] {0.5}
		    \psfrag {g} [cc][cc][1][0] 
		    	{\hspace{4em}$\gamma \brac{\vthi/\lref}$}
		    \psfrag {v} [cc][cc][1][0] {}
		    \psfrag {th} [cc][cc][1][0] {}
		    \psfrag {} [cc][cc][1][0] {}
		    \psfrag {/L} [cc][cc][1][0] {}
		    \psfrag {ref} [cc][cc][1][0] {}
			\label{fig:Lin:Gamma:Q}
			\includegraphics[width=0.5\columnwidth]
		 	{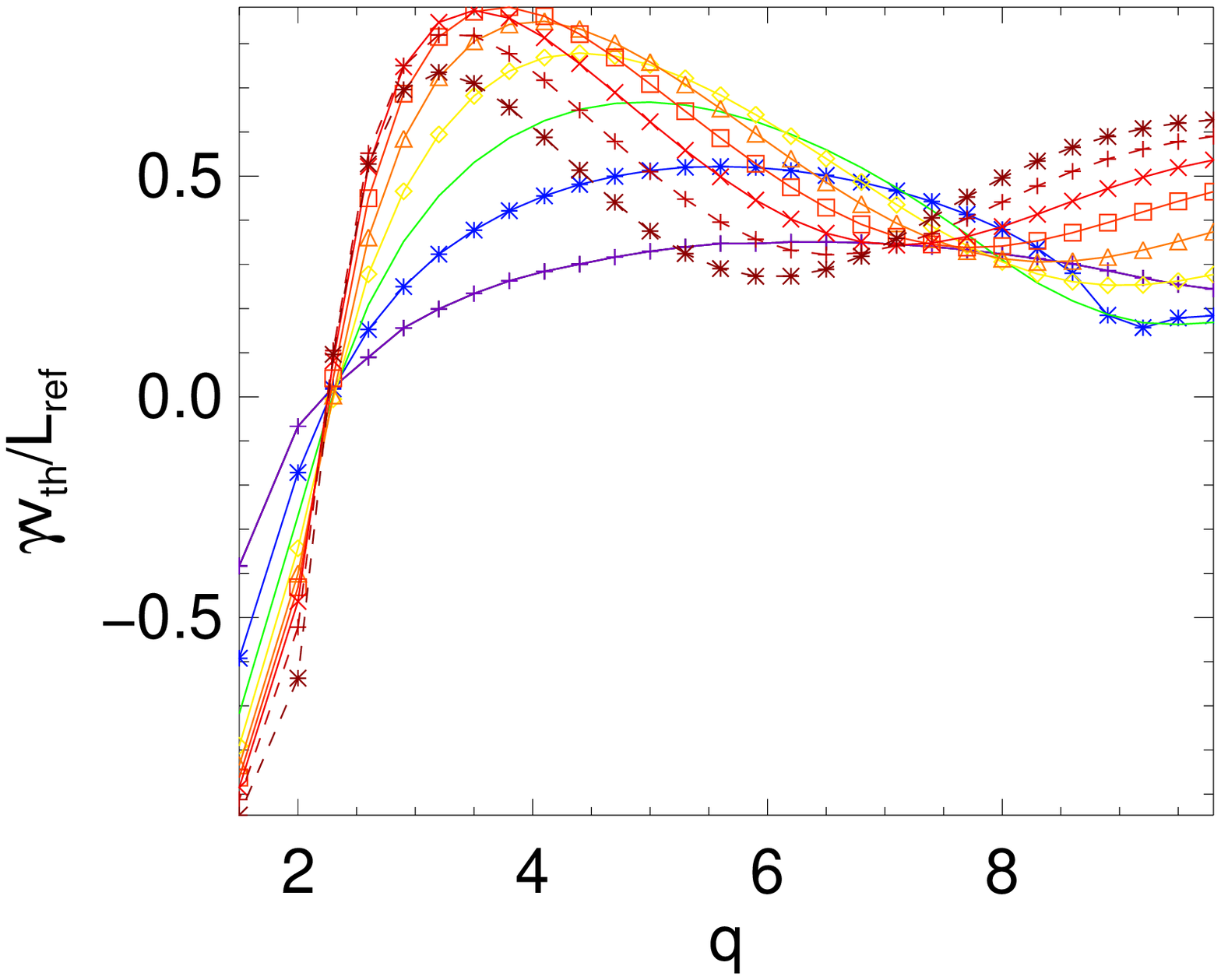}
		}
        \hspace{0.001\columnwidth}
    	\subfigure[]{
    	    \psfrag {4} [cc][cc][0.9][0] {4}
		    \psfrag {5} [cc][cc][0.9][0] {5}
		    \psfrag {6} [cc][cc][0.9][0] {6}
		    \psfrag {7} [cc][cc][0.9][0] {7}
		    \psfrag {8} [cc][cc][0.9][0] {8}
		    \psfrag {9} [cc][cc][0.9][0] {9}
		    \psfrag {shear} [cc][cc][1][0] 
		    	{\raisebox{-2.5ex}{$\shat$}}
		    \psfrag {-1.0} [cc][cc][0.9][0] {-1.0}
		    \psfrag {-0.5} [cc][cc][0.9][0] {-0.5}
		    \psfrag {0.0} [cc][cc][0.9][0] {0.0}
		    \psfrag {0.5} [cc][cc][0.9][0] {0.5}
		    \psfrag {1.0} [cc][cc][0.9][0] {1.0}
		    \psfrag {g} [cc][cc][1][0] 
		    	{\hspace{4em}$\gamma \brac{\vthi/\lref}$}
		    \psfrag {v} [cc][cc][1][0] {}
		    \psfrag {th} [cc][cc][1][0] {}
		    \psfrag {} [cc][cc][1][0] {}
		    \psfrag {/L} [cc][cc][1][0] {}
		    \psfrag {ref} [cc][cc][1][0] {}			
		    \label{fig:Lin:Gamma:Shat}
			\raisebox{0.5ex}
			{\includegraphics[width=0.5\columnwidth]
		 	{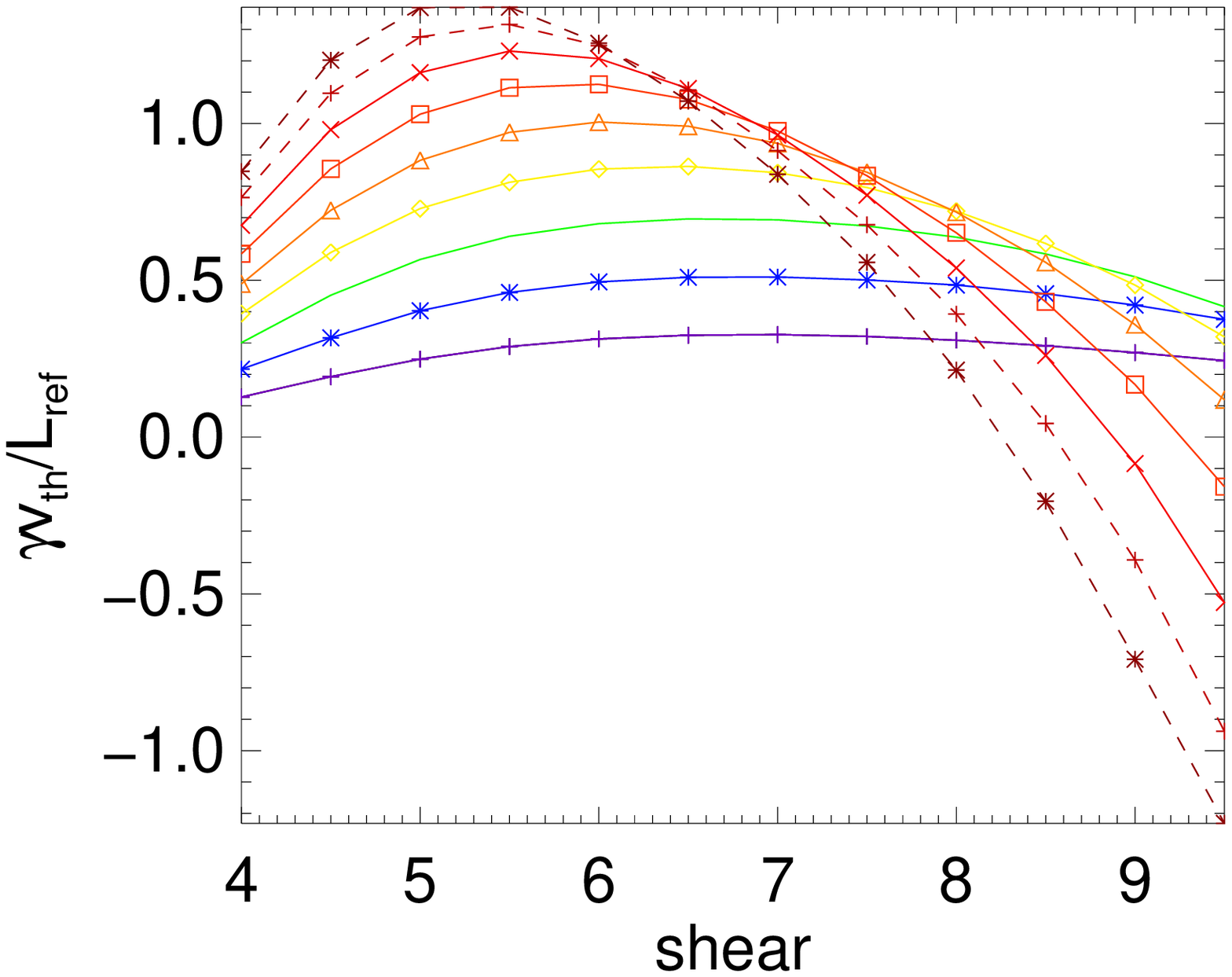}}
		}
	}
    \caption[Text]{
    $\gmtm$ as a function of (a) $q$ and (b) $\shat$, for each of the \ky\rhoi{} values in the key of \fig{fig:Lin:Spectra:BetaAndColl}.
    }
    \label{fig:Lin:Spectra:Q}
\end{figure}

%
The safety factor, $q$, and magnetic shear, $\shat$, are significantly larger in the edge plateau of MAST than at the mid-radius surface studied in \Ref{Applegate2007}.
\Fig{fig:Lin:Gamma:Q} illustrates, for a range of \ky{} values, the complicated dependence of $\gmtm$ on $q$.
$\gmtm$ exhibits multiple peaks at different $q$ locations that vary with \ky{}: e.g. the \ky\rhoi{}=1 mode has growth rate peaks at $q\sim3$ and $q\sim9$, and a local minimum at $q\sim6$.
Interestingly the MTM is stable for $q\lesssim2$.
These features may be related to the impact of $q$ on bounce and transit frequencies, which are inversely proportional to $q$.
\Fig{fig:Lin:Gamma:Shat} shows results from a scan in $\shat$, and clearly indicates that each mode has a preferred value of $\shat$ that maximises $\gmtm$.
The most unstable $\shat$ decreases as \ky{} increases, and at lower $\shat$ the peak of the $\gmtm$ spectrum moves to higher \ky\rhoi{}.
Changes in $\shat$ affect the magnetic drift frequency, $\omega_D$, which in the next section will be shown to impact on the growth rate.

\subsection{Drift frequency}
%
\begin{figure}[htb]
    \centering
    \resizebox{0.5\columnwidth}{!}{
	   	    \psfrag {0.5} [cc][cc][0.9][0] {0.5 }
		    \psfrag {1.0} [cc][cc][0.9][0] {1.0 }
		    \psfrag {1.5} [cc][cc][0.9][0] {1.5 }
		    \psfrag {2.0} [cc][cc][0.9][0] {2.0 }
		    \psfrag {2.5} [cc][cc][0.9][0] {2.5 }
		    \psfrag {e} [cc][cc][1][0] 
		    	{\raisebox{-2.5ex}{$\epsilon_l$}}
		    \psfrag {l} [cc][cc][1][0] {}
		    \psfrag {-1.5} [cc][cc][0.9][0] {-1.5 }
		    \psfrag {-1.0} [cc][cc][0.9][0] {-1.0 }
		    \psfrag {-0.5} [cc][cc][0.9][0] {-0.5 }
		    \psfrag {0.0} [cc][cc][0.9][0] {0.0 }
		    \psfrag {g} [cc][lt][1][0] 
		    	{\hspace{2em}$\gamma \brac{\vthi/\lref}$}
		    \psfrag {v} [cc][cc][1][0] {}
		    \psfrag {th} [cc][cc][1][0] {}
		    \psfrag {} [cc][cc][1][0] {}
		    \psfrag {/L} [cc][cc][1][0] {}
		    \psfrag {ref} [cc][cc][1][0] {}
			\includegraphics[width=0.5\columnwidth]
		 	{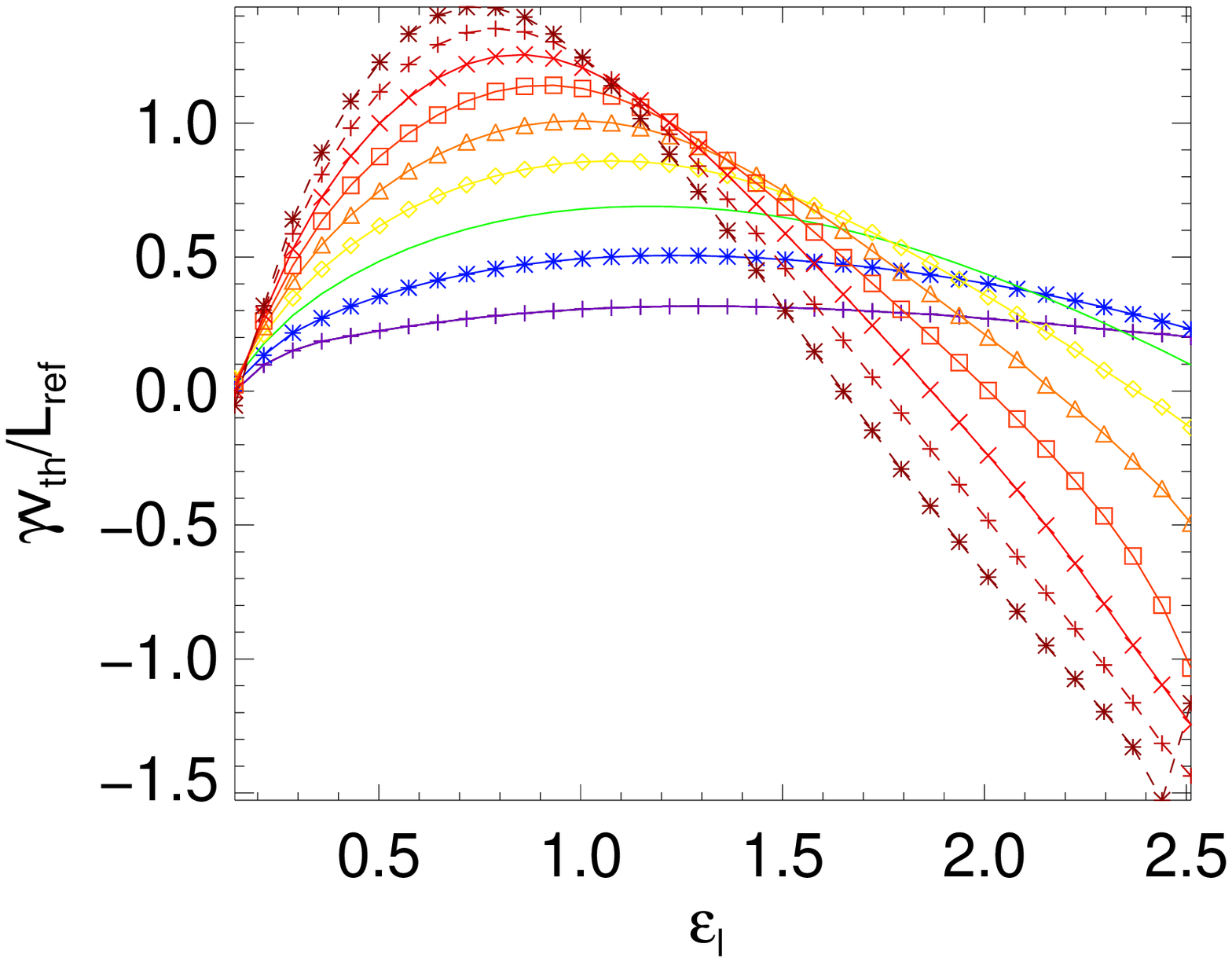}
	}
    \caption[Text]{
    $\gmtm$ as a function of magnetic drift strength factor, $\epsilon_l$, for each of the \ky\rhoi{} values in the key of \fig{fig:Lin:Spectra:BetaAndColl}.
    }
 	\label{fig:Lin:Gamma:Epsl}
\end{figure}
%
In the simple \sa{} model used here the curvature and $\vect{\nabla}B$ drifts have an equal velocity independent factor, $D(\theta)$, given at zero ballooning angle ($\theta_0=0$) by:
\begin{equation}
	D\brac{\theta}=
		\epsilon_l\left[
			\cos\brac{\theta}
			-\left(
				\alpha\sin\brac{\theta}
				-\shat\theta
			\right)\sin\brac{\theta}
		\right]
\label{eq:Lin:DriftTerm}
\end{equation}
and combine to give the magnetic drift frequency, $\omega_D\propto \ky  D\brac{\theta} \brac{\vpar^2+\vperp^2/2}$.
A scan in drift frequency was performed by varying $\epsilon_l$ around its reference value, $\epsilon_l=1.435$, with all other parameters fixed.
\Fig{fig:Lin:Gamma:Epsl} shows that the growth rate peaks at a particular $\epsilon_l$, which varies with \ky{}, and that $\gmtm$ is more sensitive to the drifts at higher \ky{}.
The peak growth rate occurs at a drift strength factor that decreases approximately linearly with \ky{}, suggesting an optimal value of $\omega_D$ for which $\gmtm$ of each mode is maximised.
This indicates that some form of drift resonance may be important.
These MTMs are stable in the absence of magnetic drifts (i.e. $\epsilon_l=0$), showing that the slab drive is insufficient for instability \footnotemark{}.
\footnotetext{In \Ref{Applegate2007} a residual instability remained in the absence of magnetic drifts (provided $\phi$ was retained), which may be due to a stronger drive from more passing particles at lower $\epsilon$.}

Independent scans in the magnetic drift frequencies for trapped and passing particles, reveal that $\gmtm$ is most sensitive to the trapped particle drifts and that passing particle drifts are unimportant.
We note that the magnetic shear scan of \fig{fig:Lin:Gamma:Shat} was effectively a scan in the radial component of $\omega_D$, which is $\propto \sin\brac{\theta}\shat\theta$ from \eqn{eq:Lin:DriftTerm}.
Therefore the similarity of \figs{fig:Lin:Gamma:Shat} and \ref{fig:Lin:Gamma:Epsl} indicates that the radial component of the magnetic drift is the dominant influence on the drive mechanism. 
In ballooning space the radial wavenumber exceeds \ky{} for $\shat \theta > 1$, which arises for $\theta>0.13$ in the edge, and for $\theta>3.49$ for the mid-radius MAST parameters of \Ref{Applegate2007}.
The radial component of the drift frequency for trapped particles is clearly more significant at the edge of MAST than at mid-radius. 
Trapped particles and their radial drifts seem to play an essential role in the MTM drive mechanism at large $\epsilon$.
Analytic theories of the MTM either neglect the magnetic drift frequency, $\omega_D$, or assume $\omega_D \ll \omega$.
The magnetic drifts can neither be neglected nor treated as small for these edge MTMs.

\subsection{Frequencies}
%
\begin{figure}[htb]
    \centering
    \resizebox{1.0\columnwidth}{!}{
    	\subfigure[]{
			\psfrag {1} [cc][cc][0.9][0] {1}
		    \psfrag {2} [cc][cc][0.9][0] {2}
		    \psfrag {3} [cc][cc][0.9][0] {3}
		    \psfrag {4} [cc][cc][0.9][0] {4}
		    \psfrag {v} [cc][cc][1][0] {}
		    \psfrag {perp} [cc][cc][1][0] 
		    	{\raisebox{1ex}{\hspace{1em}\vperp/\vthe{}}}
		    \psfrag {/v} [cc][cc][1][0] {}
		    \psfrag {th,e} [cc][cc][1][0] {}
		    \psfrag {0} [cc][cc][0.9][0] {}
		    \psfrag {par} [cc][cc][1][0] 
		    	{\raisebox{2ex}{\hspace{1em}\vpar/\vthe{}}}
		    \psfrag {gp-gm} [cc][cc][1][0]
		    	{\raisebox{-1ex}
		    		{\hspace{1em}$\left|g_+-g_-\right|$}}
	        \psfrag {0.00} [cc][cc][0.9][0] {}
		    \psfrag {0.20} [cc][cc][0.9][0] {0.2}
		    \psfrag {0.40} [cc][cc][0.9][0] {}
		    \psfrag {0.60} [cc][cc][0.9][0] {0.6}
		    \psfrag {0.80} [cc][cc][0.9][0] {}
		    \psfrag {1.00} [cc][cc][0.9][0] {1.0}
		    \label{fig:Lin:Freq:Res:Freqs}
			\includegraphics[width=0.31\columnwidth]
			{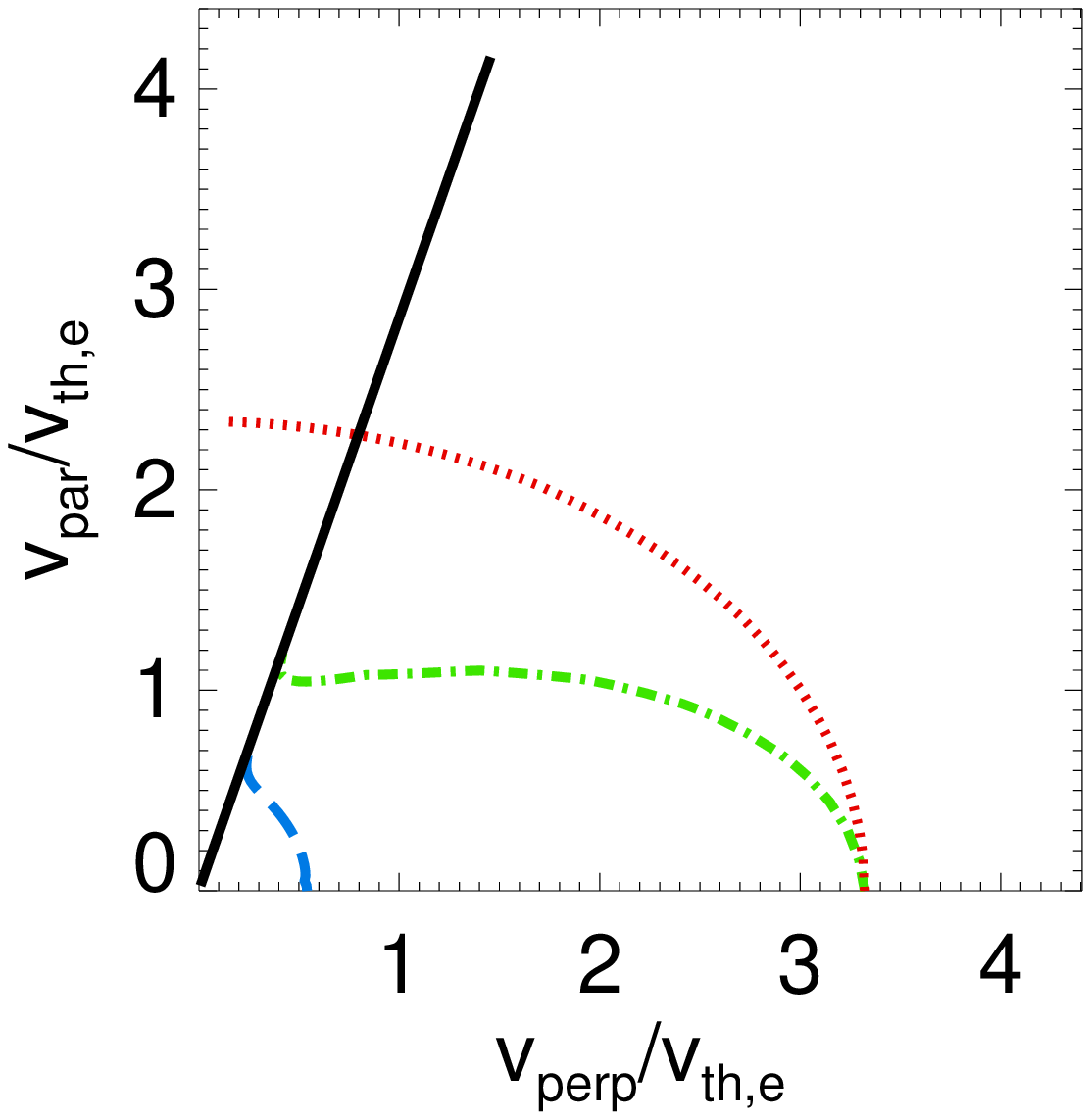}
		}
		\subfigure[]{
			\psfrag {1} [cc][cc][0.9][0] {1}
		    \psfrag {2} [cc][cc][0.9][0] {2}
		    \psfrag {3} [cc][cc][0.9][0] {3}
		    \psfrag {4} [cc][cc][0.9][0] {4}
		    \psfrag {v} [cc][cc][1][0] {}
		    \psfrag {perp} [cc][cc][1][0] 
		    	{\raisebox{1ex}{\hspace{1em}\vperp/\vthe{}}}
		    \psfrag {/v} [cc][cc][1][0] {}
		    \psfrag {th,e} [cc][cc][1][0] {}
		    \psfrag {0} [cc][cc][0.9][0] {}
		    \psfrag {par} [cc][cc][1][0] 
		    	{\raisebox{2ex}{\hspace{1em}\vpar/\vthe{}}}
		    \psfrag {gp-gm} [cc][cc][1][0]
		    	{\raisebox{-1ex}
		    		{\hspace{1em}$\left|g_+-g_-\right|$}}
	        \psfrag {0.00} [cc][cc][0.9][0] {}
		    \psfrag {0.20} [cc][cc][0.9][0] {0.2}
		    \psfrag {0.40} [cc][cc][0.9][0] {}
		    \psfrag {0.60} [cc][cc][0.9][0] {0.6}
		    \psfrag {0.80} [cc][cc][0.9][0] {}
		    \psfrag {1.00} [cc][cc][0.9][0] {1.0}
		    \label{fig:Lin:Freq:Res:Vnewk}
	    	\includegraphics[width=0.31\columnwidth]
			 {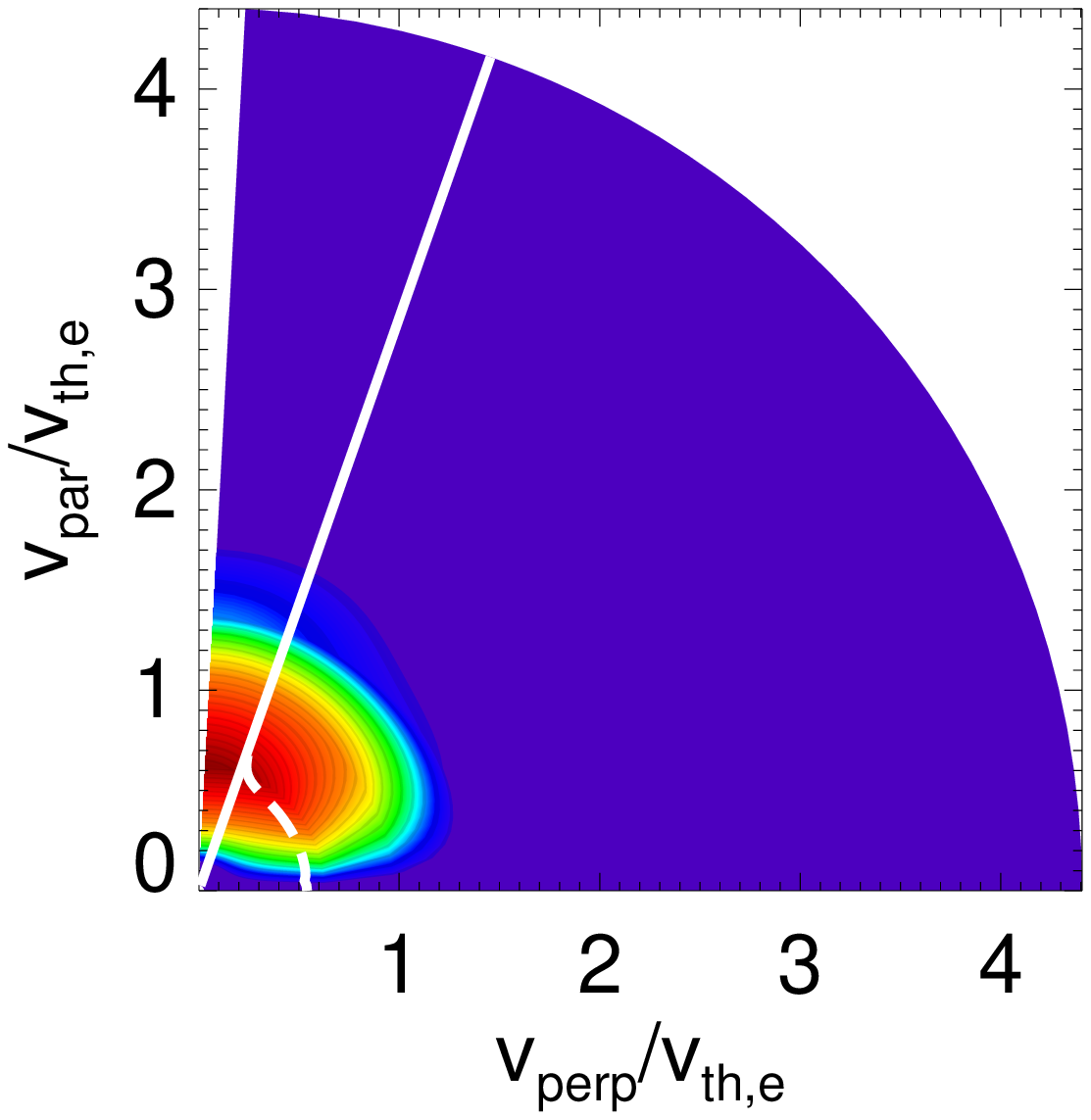}
		}
		\subfigure[]{
			\psfrag {1} [cc][cc][0.9][0] {1}
		    \psfrag {2} [cc][cc][0.9][0] {2}
		    \psfrag {3} [cc][cc][0.9][0] {3}
		    \psfrag {4} [cc][cc][0.9][0] {4}
		    \psfrag {v} [cc][cc][1][0] {}
		    \psfrag {perp} [cc][cc][1][0] 
		    	{\raisebox{1ex}{\hspace{1em}\vperp/\vthe{}}}
		    \psfrag {/v} [cc][cc][1][0] {}
		    \psfrag {th,e} [cc][cc][1][0] {}
		    \psfrag {0} [cc][cc][0.9][0] {}
		    \psfrag {par} [cc][cc][1][0] 
		    	{\raisebox{2ex}{\hspace{1em}\vpar/\vthe{}}}
		    \psfrag {gp-gm} [cc][cc][1][0]
		    	{\raisebox{-1ex}
		    		{\hspace{1em}$\left|g_+-g_-\right|$}}
	        \psfrag {0.00} [cc][cc][0.9][0] {}
		    \psfrag {0.20} [cc][cc][0.9][0] {0.2}
		    \psfrag {0.40} [cc][cc][0.9][0] {}
		    \psfrag {0.60} [cc][cc][0.9][0] {0.6}
		    \psfrag {0.80} [cc][cc][0.9][0] {}
		    \psfrag {1.00} [cc][cc][0.9][0] {1.0}
		    \label{fig:Lin:Freq:Res:NoVnewk}
	    	\includegraphics[width=0.45\columnwidth]
	 			{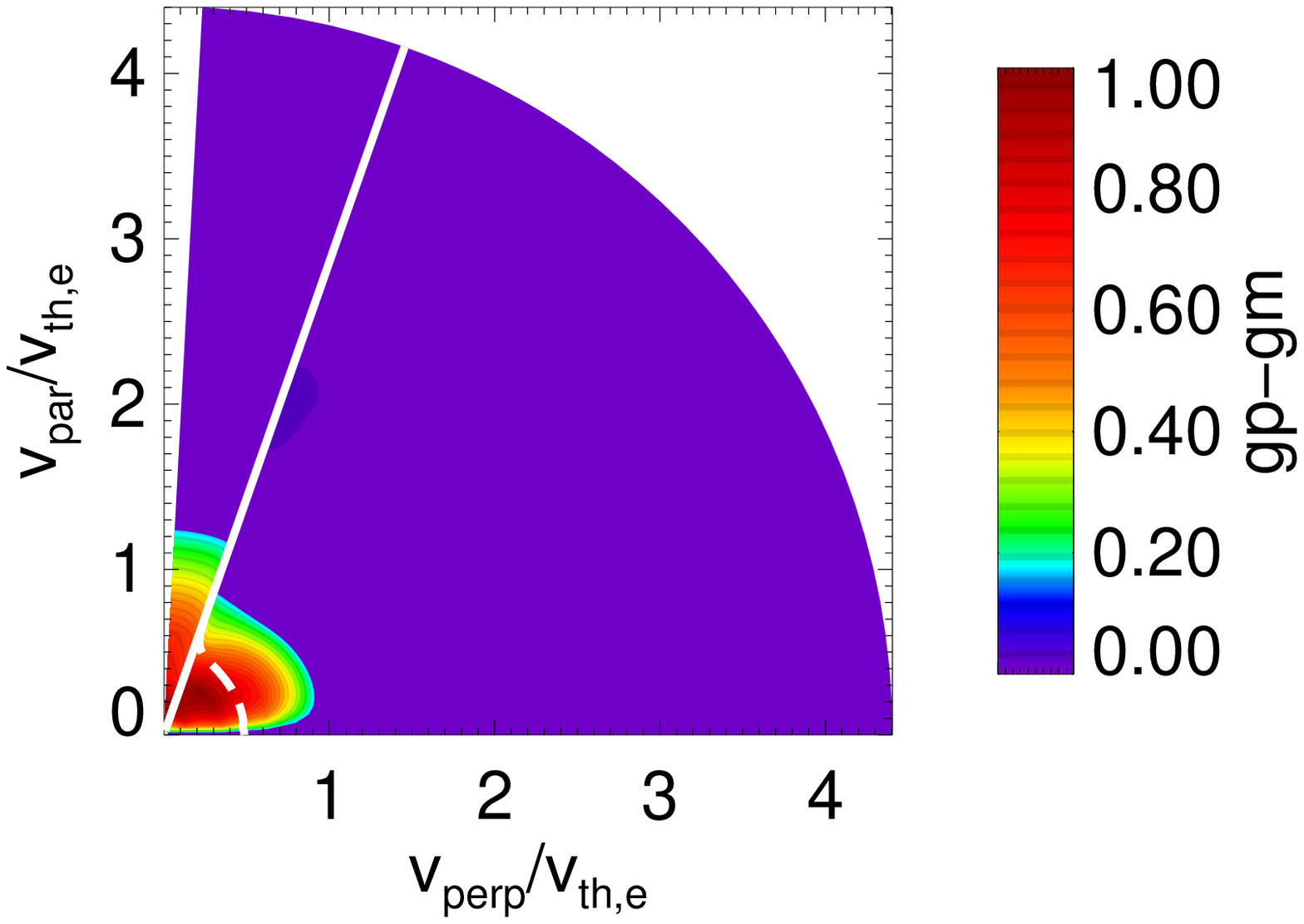}
		}
	}
    \caption[Text]{
    (a) The location in velocity space where $\omega$ for $\ky\rhoi=0.6$ matches the drift (evaluated at $\theta=0$, {\color{red} \dotted}), precession ({\color{green} \chain}) and bounce ({\color{blue} \dashed}) frequencies. 
    The current carrying asymmetry in the perturbed electron distribution function, $\dg{}=\left|g_+-g_-\right|$ for (b) \nue=1.975 and (c) \nue=0.0.
    In (b) and (c) the trapped passing boundary and bounce resonance are indicated as a solid straight line (\full) and dashed curve (\dashed), respectively.
    }
    \label{fig:Lin:Freq:Res}
\end{figure}

%
It is of direct interest to analytic theory to ask how the MTM mode frequency, $\omega$, compares with the natural electron orbit frequencies: the bounce frequency, $\omega_b$, the drift frequency at $\theta=0$, $\omega_D$, and the precession frequency, $\omega_p=\left<\omega_D\right>$.
\Fig{fig:Lin:Freq:Res:Freqs} shows where in velocity space each of these frequencies (which depend on \vpar{}, \vperp{} and \ky{}) matches the absolute mode frequency for the \ky\rhoi{}=0.6 mode.
The contours indicate that for a thermal electron $\omega_D$, $\omega_b$, $\omega_p \sim \order{\omega}$.
All three resonances lie within the range $0.5\vthe-3.5\vthe$, and may therefore have significant impact.

This poses several thoughts for analytic theory.
Firstly the perturbation changes significantly in one bounce period due to the proximity of $\omega$ and $\omega_b$.
Bounce averaging, which is often used to simplify the trapped particle response, is therefore not appropriate here.
Secondly the magnetic drift frequencies are of the same order as the mode frequency, and cannot be treated as small.

The current carrying asymmetry in the perturbed electron distribution function, \dg{}, can be obtained from the non-adiabatic perturbed electron distribution function, $g$, via:
\begin{equation}
\dg = 
	\left| 
		g(E,\mu,+)- g(E,\mu,-)
	\right|
\label{eq:Lin:DGDef}
\end{equation} 
where the arguments are energy, $E$, magnetic moment, $\mu$, and $\sgn\brac{\vpar}$.
\Figs{fig:Lin:Freq:Res:Vnewk} and \ref{fig:Lin:Freq:Res:NoVnewk} show \dg{} normalised and evaluated at $\theta=0$ for MTM simulations respectively with and without collisions.
Both plots indicate that the trapped electrons carry current.
\dg{} has clear peaks near the bounce/transit resonance, and significant amplitude around the thermal velocity.
The discontinuity in \dg{} at the trapped-passing boundary in the absence of collisions is smoothed on including collisions.

\subsection{Impact of electron FLR effects}
\begin{figure}[htb]
    \centering
    \resizebox{0.5\columnwidth}{!}{
    	\psfrag {0.2} [cc][cc][0.9][0] {0.2 }
	    \psfrag {0.4} [cc][cc][0.9][0] {0.4 }
	    \psfrag {0.6} [cc][cc][0.9][0] {0.6 }
	    \psfrag {0.8} [cc][cc][0.9][0] {0.8 }
	    \psfrag {1.0} [cc][cc][0.9][0] {1.0 }
	    \psfrag {k} [cc][cc][1][0] 
   			{\raisebox{-2.5ex}{\hspace{1em}\ky$\rho_i$}}
	    \psfrag {y} [cc][cc][1][0] {}
    	\psfrag {r} [cc][cc][1][0] {}
    	\psfrag {i} [cc][cc][1][0] {}
    	\psfrag {0.0} [cc][cc][0.9][0] {0.0 }
    	\psfrag {g} [cc][cc][1][0] 
	    	{\raisebox{2ex}{\hspace{2em}$\gamma \brac{\vthi/\lref}$}}
	    \psfrag {v} [cc][cc][1][0] {}
	    \psfrag {th} [cc][cc][1][0] {}
    	\psfrag {/L} [cc][cc][1][0] {}
    	\psfrag {ref} [cc][cc][1][0] {}
    	\label{fig:Lin:FLR:A}
			\includegraphics[width=0.5\columnwidth]
		 	{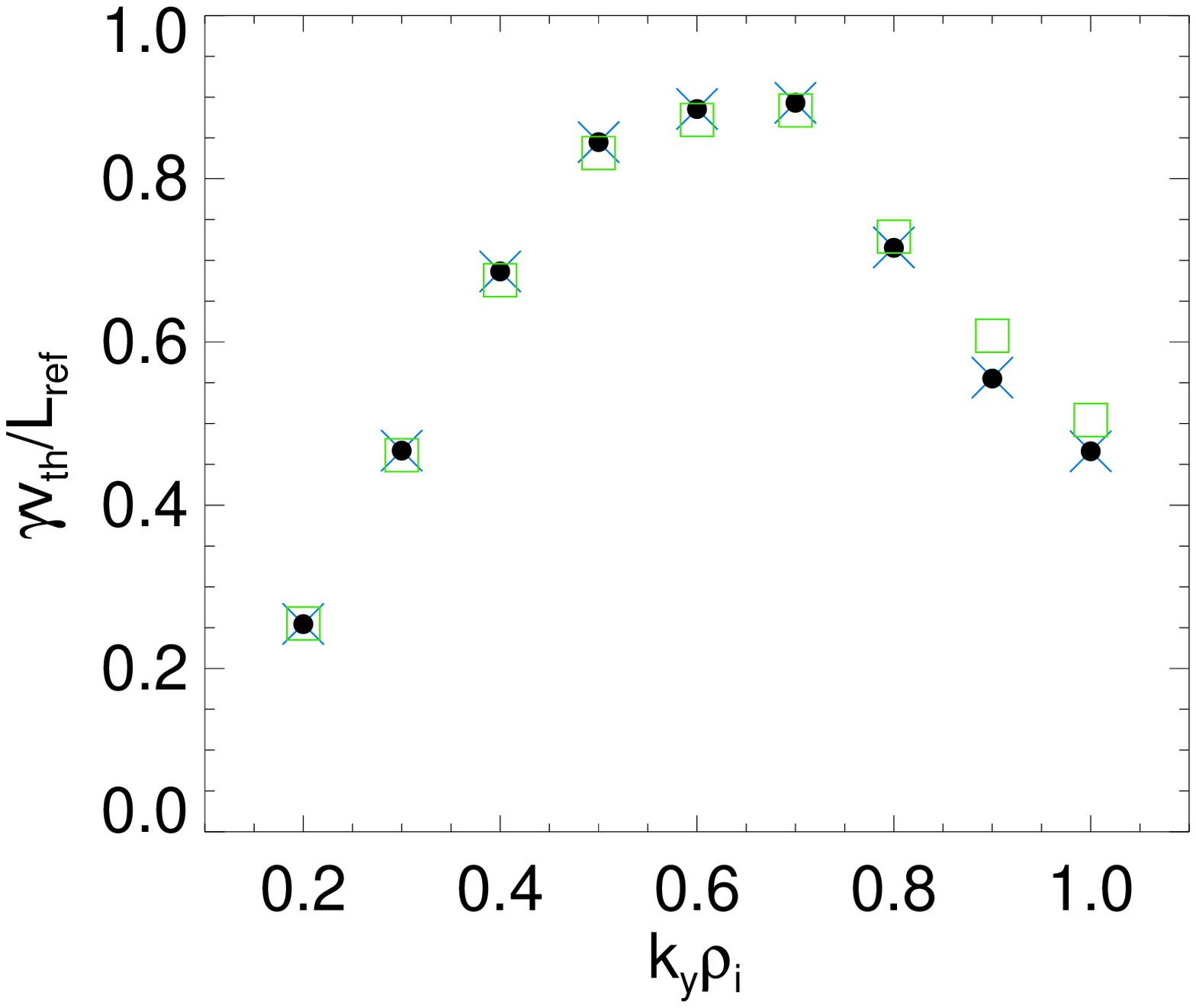}
	}
    \caption[Text]{
    \gmtm{} as a function of \ky\rhoi{} for Bessel function arguments multiplied by 0.1 ({\color{black} \fullcircle{}}), 1 ({\color{blue} $\times$}) and 10.0 ({\color{green} \opensquare{}}).
    }
    \label{fig:Lin:Spectra:FLR}
\end{figure}

Whilst the characteristic binormal wavenumbers associated with these MTMs satisfy $\ky \rhoe\ll1$, higher values of the radial wavenumber, $\kx \rhoe\sim\order{1}$, are needed to describe the $\phi$ eigenfunction at high $\theta$.
We have assessed the importance of electron FLR effects, which enter the linear drive terms of the gyrokinetic equation via Bessel functions, by repeating MTM simulations (with adiabatic ions and in the absence of collisions) with the Bessel function arguments multiplied by a factor $\bfac=(0.1, 1.0, 10.0)$.
The $\bfac=0.1$ simulations produced eigenfunctions that were practically identical to those with $\bfac=1.0$, and \fig{fig:Lin:Spectra:FLR} shows a negligible impact on the growth rate spectra.
This suggests that electron FLR effects are not important for these MTMs.

%
%

\section{Conclusions}
\label{sec:Conclusion}
Gyrokinetic simulations have found that microtearing modes (MTMs) are unstable in both STs and 
large aspect ratio devices.
Recent simulations find that MTMs are also unstable in the shallow gradient region just inboard of the MAST H-mode pedestal, which may impact on its evolution between ELMs.
Analytic theory has proposed two different linear drive mechanisms for MTMs: one based on a simple slab model, and the other requiring trapped particles.
Both mechanisms require a finite rate of electron-ion collisions ($\nue>0$) for instability. 

A detailed study of the basic linear properties of edge MTMs has been performed using a simplified circular \sa{} model fit to the local equilibrium at the edge of MAST.
Consistent with existing MTM theories it is found that the mode frequency $\omega\sim\wde$, and that the modes are unstable only if finite stability thresholds are exceeded in $dT_e/dr$ and $\beta_e$.
The growth rate's dependence on $\nue$, however, is in conflict with existing analytic models.
In both the \sa{} model equilibrium and the fully shaped MAST edge equilibrium, it is found that $\gmtm$ for the dominant mode is maximised in the absence of collisions (i.e. at $\nue=0$), where the existing drive mechanisms should vanish.
Trapped particles are essential to drive these MTMs, and sensitivity of $\gmtm$ to the magnetic drift frequency suggests that a drift resonance may be involved.
$\gmtm$ rises with the trapped particle fraction, and the $\ky\rhoi$ associated with the dominant mode drops with increasing $\shat$. 
The mode frequency and the thermal trapped electron bounce, precession and drift frequencies are all of the same order. 
To our knowledge this regime has not been addressed by an existing analytic theory.
In present models the magnetic drifts are typically neglected or treated as small, and any trapped particle response is usually obtained using bounce averaging.
Neither of these approximations are valid here.

The drive for similar MTMs, at $\ky\rhoi\sim\order{1}$, should be enhanced in the high magnetic shear region of the edge plateau in tokamaks, and perhaps especially in STs.
Similar MTMs have also recently been found unstable towards the edge of conventional aspect ratio tokamaks including JET \cite{Saarelma2012,SaarelmaIAEA2012} and ASDEX Upgrade \cite{Told2008}, suggesting that this drive mechanism may have wide ranging significance.

\appendix
\section{Sensitivity of results to grid resolutions}
\label{sec:NumDiss}
\begin{figure}[htb]
    \centering
    \resizebox{1.0\columnwidth}{!}{
    	\subfigure[]{
    		\large
    		\psfrag {1x10} [cc][cc][0.8][0] {1$\times$10}
		    \psfrag {-1} [cc][cc][0.6][0] {\hspace{0.6em}-1}
		    \psfrag {0} [cc][cc][0.6][0] {\hspace{0.6em}0}
		    \psfrag {1} [cc][cc][0.6][0] {\hspace{0.6em}1}
		    \psfrag {2} [cc][cc][0.6][0] {\hspace{0.6em}2}
		    \psfrag {n} [cc][cc][1][0] 
		    	{\raisebox{-2.5ex}
		    		{\hspace{2em}$\nue \brac{\vthi/\lref}$}}
		    \psfrag {ei} [cc][cc][1][0] {}
		    \psfrag {v} [cc][cc][1][0] {}
		    \psfrag {th} [cc][cc][1][0] {}
		    \psfrag {/L} [cc][cc][1][0] {}
		    \psfrag {ref} [cc][cc][1][0] {}
		    \psfrag {} [cc][cc][1][0] {}
		    \psfrag {0.0} [cc][cc][0.9][0] {0.0}
		    \psfrag {0.2} [cc][cc][0.9][0] {0.2}
		    \psfrag {0.4} [cc][cc][0.9][0] {0.4}
		    \psfrag {0.6} [cc][cc][0.9][0] {0.6}
		    \psfrag {0.8} [cc][cc][0.9][0] {0.8}
		    \psfrag {1.0} [cc][cc][0.9][0] {1.0}
		    \psfrag {g} [cc][cc][1][0] 
		    	{\hspace{2em}$\gamma \brac{\vthi/\lref}$}
			\label{fig:NumDisVspace:Ntheta}
			\includegraphics[width=0.5\columnwidth]
		 		{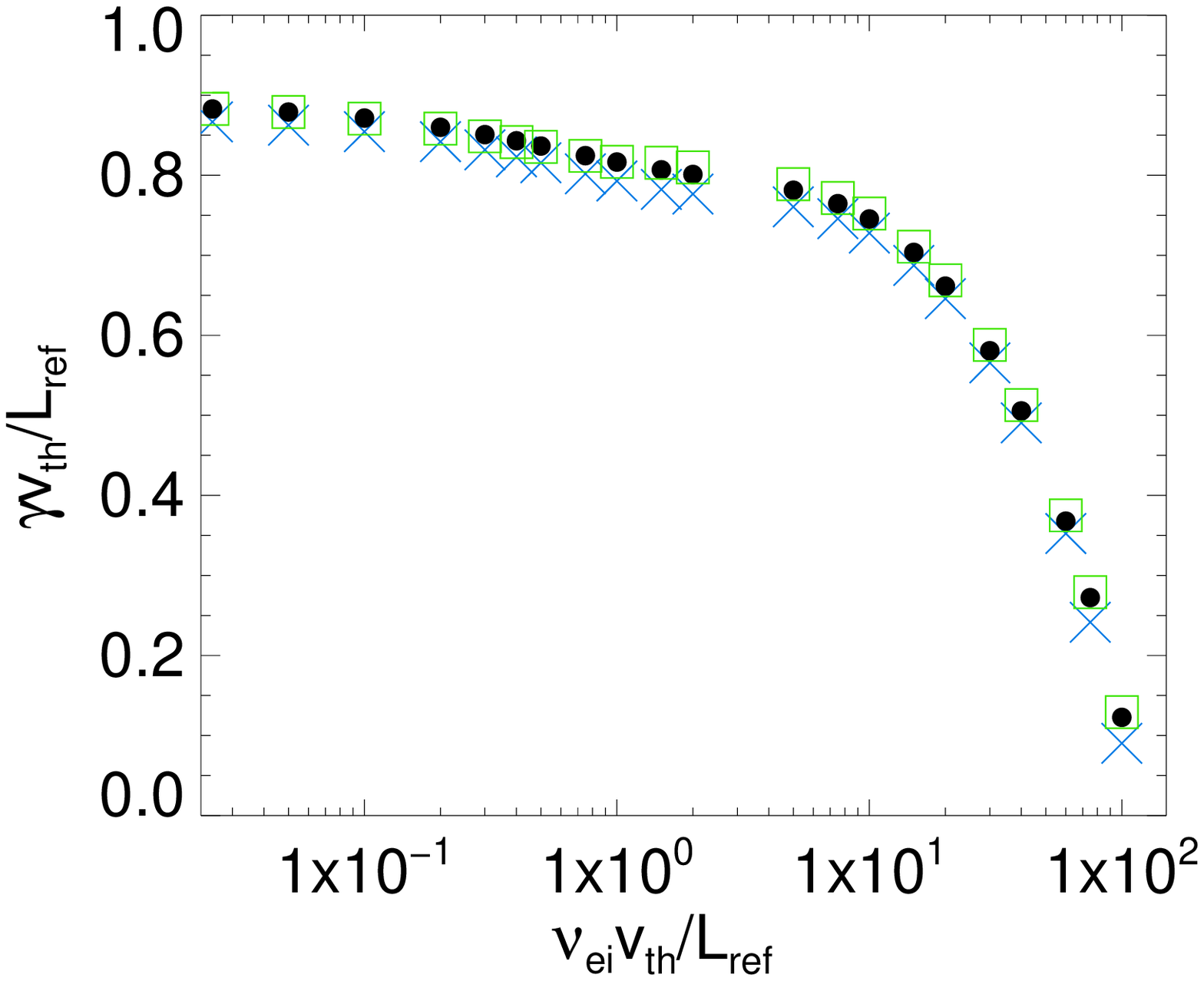}
		}
		\subfigure[]{
			\large
			\psfrag {1x10} [cc][cc][0.8][0] {1$\times$10}
		    \psfrag {-1} [cc][cc][0.6][0] {\hspace{0.6em}-1}
		    \psfrag {0} [cc][cc][0.6][0] {\hspace{0.6em}0}
		    \psfrag {1} [cc][cc][0.6][0] {\hspace{0.6em}1}
		    \psfrag {2} [cc][cc][0.6][0] {\hspace{0.6em}2}
		    \psfrag {n} [cc][cc][1][0] 
		    	{\raisebox{-2.5ex}
		    		{\hspace{2em}$\nue \brac{\vthi/\lref}$}}
		    \psfrag {ei} [cc][cc][1][0] {}
		    \psfrag {v} [cc][cc][1][0] {}
		    \psfrag {th} [cc][cc][1][0] {}
		    \psfrag {/L} [cc][cc][1][0] {}
		    \psfrag {ref} [cc][cc][1][0] {}
		    \psfrag {} [cc][cc][1][0] {}
		    \psfrag {0.0} [cc][cc][0.9][0] {0.0}
		    \psfrag {0.2} [cc][cc][0.9][0] {0.2}
		    \psfrag {0.4} [cc][cc][0.9][0] {0.4}
		    \psfrag {0.6} [cc][cc][0.9][0] {0.6}
		    \psfrag {0.8} [cc][cc][0.9][0] {0.8}
		    \psfrag {1.0} [cc][cc][0.9][0] {1.0}
		    \psfrag {g} [cc][cc][1][0] 
		    	{\hspace{2em}$\gamma \brac{\vthi/\lref}$}
			\label{fig:NumDisVspace:Ngauss}
			\includegraphics[width=0.5\columnwidth]
		 		{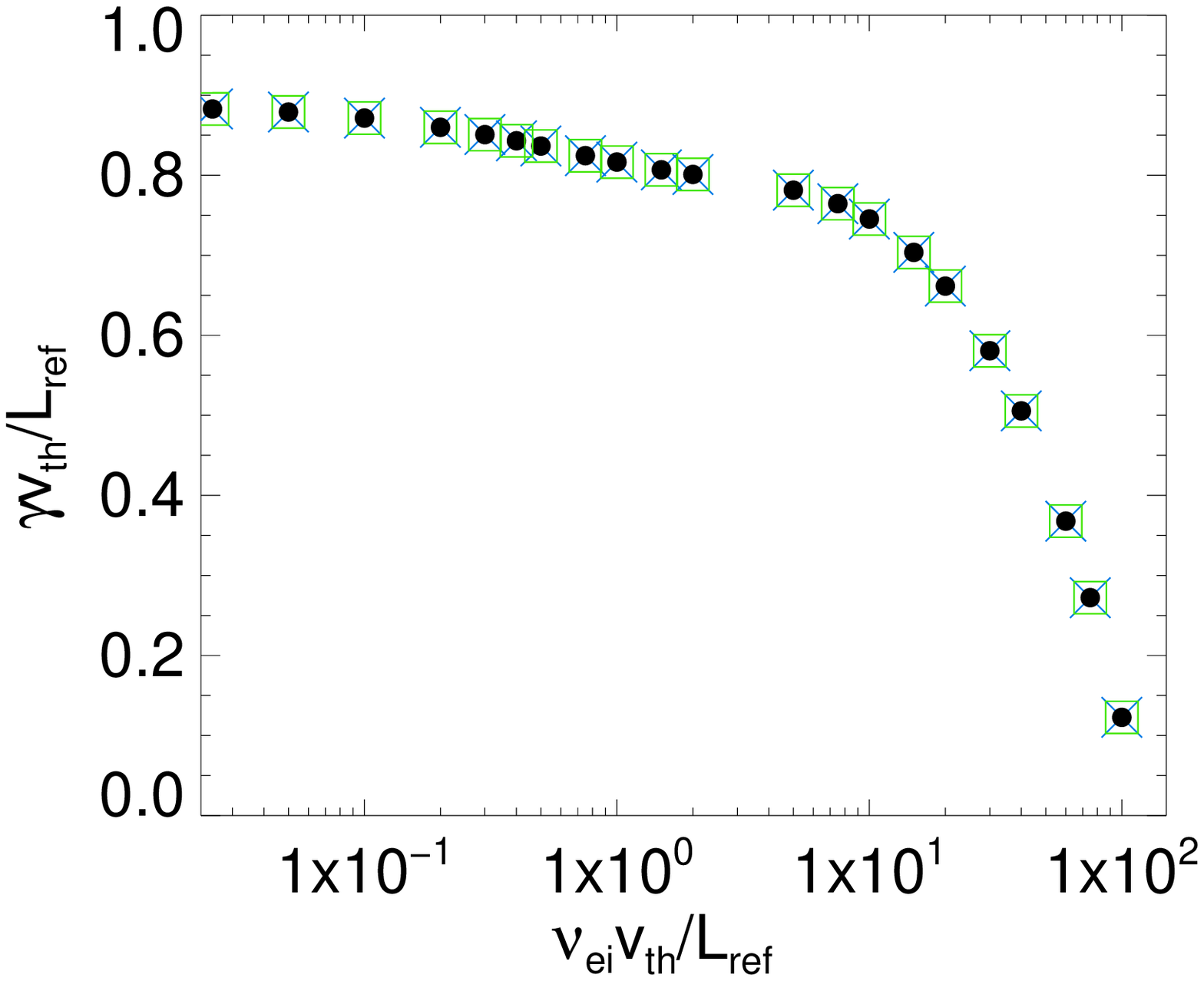}

		}
		\subfigure[]{
			\large
			\psfrag {1x10} [cc][cc][0.8][0] {1$\times$10}
		    \psfrag {-1} [cc][cc][0.6][0] {\hspace{0.6em}-1}
		    \psfrag {0} [cc][cc][0.6][0] {\hspace{0.6em}0}
		    \psfrag {1} [cc][cc][0.6][0] {\hspace{0.6em}1}
		    \psfrag {2} [cc][cc][0.6][0] {\hspace{0.6em}2}
		    \psfrag {n} [cc][cc][1][0] 
		    	{\raisebox{-2.5ex}
		    		{\hspace{2em}$\nue \brac{\vthi/\lref}$}}
		    \psfrag {ei} [cc][cc][1][0] {}
		    \psfrag {v} [cc][cc][1][0] {}
		    \psfrag {th} [cc][cc][1][0] {}
		    \psfrag {/L} [cc][cc][1][0] {}
		    \psfrag {ref} [cc][cc][1][0] {}
		    \psfrag {} [cc][cc][1][0] {}
		    \psfrag {0.0} [cc][cc][0.9][0] {0.0}
		    \psfrag {0.2} [cc][cc][0.9][0] {0.2}
		    \psfrag {0.4} [cc][cc][0.9][0] {0.4}
		    \psfrag {0.6} [cc][cc][0.9][0] {0.6}
		    \psfrag {0.8} [cc][cc][0.9][0] {0.8}
		    \psfrag {1.0} [cc][cc][0.9][0] {1.0}
		    \psfrag {g} [cc][cc][1][0] 
		    	{\hspace{2em}$\gamma \brac{\vthi/\lref}$}
		    \label{fig:NumDisVspace:Negrid}
			\includegraphics[width=0.5\columnwidth]
				{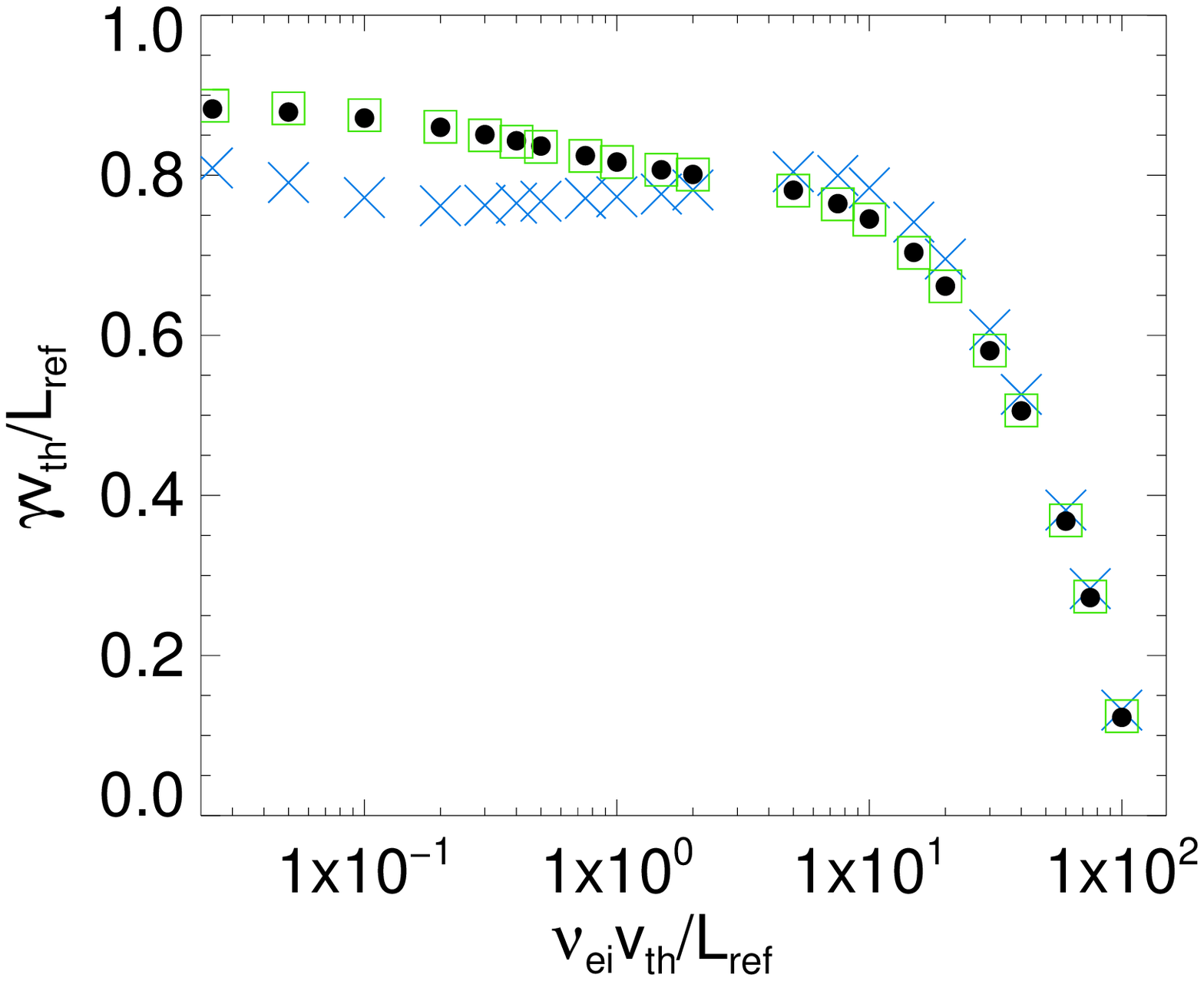}		
		}
    }
    \caption[Text]
     {
     $\gmtm$ as a function of $\nue$ for the mode with \ky\rhoi{}=0.6 for runs varying 
     (a) \codevar{ntheta} [32 ({\color{blue} $\times$}), 64 ({\color{black} \fullcircle{}}), 128 ({\color{green} \opensquare{}})] 
     (b) \codevar{ngauss} [5 ({\color{blue} $\times$}), 10 ({\color{black} \fullcircle{}}), 16 ({\color{green} \opensquare{}})] and 
     (c) \codevar{negrid} [8 ({\color{blue} $\times$}), 16 ({\color{black} \fullcircle{}}), 32 ({\color{green} \opensquare{}})].
The default grid parameters used throughout the paper are: \codevar{ntheta}=64, \codevar{ngauss}=5 and \codevar{negrid}=16.
}
    \label{fig:NumDisVspace}
\end{figure}

The smallest resolved features in numerical simulations are limited by the grid.
Collisions smooth fine scale features in velocity space, but at low collision frequency they may be insufficient to damp features at the grid scale.
Such structures may, however, become limited by diffusion arising from the numerical scheme.
If either of these unphysical grid dependent mechanisms were to influence our MTM simulations, the linear mode properties would be expected to vary with grid resolution.
\Fig{fig:NumDisVspace} demonstrates that the dependence of \gmtm{} on \nue{} (at \ky\rhoi{}=0.6) is not sensitive to increases in GS2 grid resolution parameters including: number of parallel grid points \codevar{ntheta}, which also determines the number of trapped pitch angles; number of passing pitch angles, \codevar{ngauss}; and the number of energy grid points \codevar{negrid}\footnotemark{}.
\footnotetext{See \Ref{Barnes2010b} for more details on the velocity space grid in GS2.}
This suggests that our grid resolution has little impact on the linear properties of the MTMs computed here.

%
%

Numerical dissipation is also introduced through upwinding in space and decentering in time, but has been shown to have negligible impact on our simulations.

\footnotesize
\ack
The authors wish to thank J~W~Connor and R~J~Hastie for helpful discussions.
This work was carried out using several supercomputers: HELIOS at International Fusion Energy Research Centre, Aomori, Japan, (under the Broader Approach collaboration between Euratom and Japan, implemented by Fusion for Energy and JAEA); HECToR, through EPSRC Grant No. EP/H002081/1; and HPC-FF (Forschungszentrum Juelich).
This work was partly funded by the RCUK Energy Programme under grant EP/I501045 and the European Communities under the contract of Association between EURATOM and CCFE.
The views and opinions expressed herein do not necessarily reflect those of the European Commission.

\bibliographystyle{aip}
\def\newblock{\hskip .11em plus .33em minus .07em}
\bibliography{Varenna_2012}

\begin{thebibliography}{10}

\bibitem{Furth1963}
H.~P. Furth, J.~Killeen, and M.~N. Rosenbluth,
\newblock Physics of Fluids {\bf 6}, 459 (1963).

\bibitem{Hazeltine1975}
R.~D. Hazeltine, D.~Dobrott, and T.~S. Wang,
\newblock Physics of Fluids {\bf 18}, 1778 (1975).

\bibitem{Drake1977}
J.~F. Drake and Y.~C. Lee,
\newblock Physics of Fluids {\bf 20}, 1341 (1977).

\bibitem{Hassam1980}
A.~B. Hassam,
\newblock Physics of Fluids {\bf 23}, 38 (1980).

\bibitem{Hassam1980a}
A.~B. Hassam,
\newblock Physics of Fluids {\bf 23}, 2493 (1980).

\bibitem{Catto1981a}
P.~J. Catto and M.~N. Rosenbluth,
\newblock Physics of Fluids {\bf 24}, 243 (1981).

\bibitem{Connor1990}
J.~W. Connor, S.~C. Cowley, and R.~J. Hastie,
\newblock Plasma Physics and Controlled Fusion {\bf 32}, 799 (1990).

\bibitem{Valovic2011}
M.~Valovi\v{c} et~al.,
\newblock Nuclear Fusion {\bf 51}, 073045 (2011).

\bibitem{Kaye2007a}
S.~M. Kaye et~al.,
\newblock Nuclear Fusion {\bf 47}, 499 (2007).

\bibitem{Guttenfelder2012a}
W.~Guttenfelder et~al.,
\newblock Physics of Plasmas {\bf 19}, 056119 (2012).

\bibitem{Kotschenreuther2000}
M.~Kotschenreuther et~al.,
\newblock Nuclear Fusion {\bf 40}, 677 (2000).

\bibitem{Applegate2004}
D.~J. Applegate et~al.,
\newblock Physics of Plasmas {\bf 11}, 5085 (2004).

\bibitem{Wilson2004a}
H.~R. Wilson et~al.,
\newblock Nuclear Fusion {\bf 44}, 917 (2004).

\bibitem{Roach2005}
C.~M. Roach et~al.,
\newblock Plasma Physics and Controlled Fusion {\bf 47}, B323 (2005).

\bibitem{Applegate2007}
D.~J. Applegate et~al.,
\newblock Plasma Physics and Controlled Fusion {\bf 49}, 1113 (2007).

\bibitem{Guttenfelder2012}
W.~Guttenfelder et~al.,
\newblock Physics of Plasmas {\bf 19}, 022506 (2012).

\bibitem{Told2008}
D.~Told, F.~Jenko, P.~Xanthopoulos, L.~D. Horton, and E.~Wolfrum,
\newblock Physics of Plasmas {\bf 15}, 102306 (2008).

\bibitem{Predebon2010}
I.~Predebon, F.~Sattin, M.~Veranda, D.~Bonfiglio, and S.~Cappello,
\newblock Physical Review Letters {\bf 105}, 195001 (2010).

\bibitem{Carmody2012}
D.~Carmody et~al.,
\newblock {Microtearing Mode Fluctuations in Reversed Field Pinch Plasmas},
\newblock in {\em 24th IAEA Fusion Energy Conference}, San Diego, 2012.

\bibitem{Stix1973}
T.~Stix,
\newblock Physical Review Letters {\bf 30}, 833 (1973).

\bibitem{Rechester1978}
A.~Rechester and M.~N. Rosenbluth,
\newblock Physical Review Letters {\bf 40}, 38 (1978).

\bibitem{Wong2007}
K.~Wong et~al.,
\newblock Physical Review Letters {\bf 99}, 1 (2007).

\bibitem{Doerk2011}
H.~Doerk, F.~Jenko, M.~J. Pueschel, and D.~R. Hatch,
\newblock Physical Review Letters {\bf 106}, 1 (2011).

\bibitem{Guttenfelder2011}
W.~Guttenfelder et~al.,
\newblock Physical Review Letters {\bf 106}, 1 (2011).

\bibitem{Doerk2012}
H.~Doerk et~al.,
\newblock Physics of Plasmas {\bf 19}, 055907 (2012).

\bibitem{Dickinson2011}
D.~Dickinson et~al.,
\newblock Plasma Physics and Controlled Fusion {\bf 53}, 115010 (2011).

\bibitem{Saarelma2012}
S.~Saarelma et~al.,
\newblock {Pedestal Modelling Based MHD Analyses on MAST and JET Plasmas},
\newblock in {\em 39th EPS Conference on Plasma Physics}, edited by
  S.~Ratynskaya, L.~Blomberg, and A.~Fasoli, pages 1--4, Stockholm, 2012,
  European Physical Society.

\bibitem{Kotschenreuther1995}
M.~Kotschenreuther, G.~Rewoldt, and W.~M. Tang,
\newblock Computer Physics Communications {\bf 88}, 128 (1995).

\bibitem{Dickinson2012}
D.~Dickinson et~al.,
\newblock Physical Review Letters {\bf 108}, 135002 (2012).

\bibitem{RoachIAEA2012}
C.~M. Roach et~al.,
\newblock Proc. 24th IAEA FEC, San Diego, TH/5-1, to be submitted to Nuclear
  Fusion  (2012).

\bibitem{Smith2011}
D.~R. Smith, W.~Guttenfelder, B.~P. LeBlanc, and D.~R. Mikkelsen,
\newblock Plasma Physics and Controlled Fusion {\bf 53}, 035013 (2011).

\bibitem{Connor1978}
J.~W. Connor, R.~J. Hastie, and J.~B. Taylor,
\newblock Physical Review Letters {\bf 40}, 396 (1978).

\bibitem{Cowley1986}
S.~C. Cowley, R.~M. Kulsrud, and T.~S. Hahm,
\newblock Physics of Fluids {\bf 29}, 3230 (1986).

\bibitem{Snyder2001}
P.~B. Snyder and G.~W. Hammett,
\newblock Physics of Plasmas {\bf 8}, 744 (2001).

\bibitem{Petty2008}
C.~C. Petty,
\newblock Physics of Plasmas {\bf 15}, 080501 (2008).

\bibitem{Tang1985}
W.~M. Tang, G.~Rewoldt, C.~Z. Cheng, and M.~S. Chance,
\newblock Nuclear Fusion {\bf 25}, 151 (1985).

\bibitem{Belli2010}
E.~A. Belli and J.~Candy,
\newblock Physics of Plasmas {\bf 17}, 112314 (2010).

\bibitem{Roach1995}
C.~M. Roach, J.~W. Connor, and S.~Janjua,
\newblock Plasma Physics and Controlled Fusion {\bf 37}, 679 (1995).

\bibitem{SaarelmaIAEA2012}
S.~Saarelma et~al.,
\newblock Proc. 24th IAEA FEC, San Diego, TH/P3-10, to be submitted to Nuclear
  Fusion  (2012).

\bibitem{Barnes2010b}
M.~Barnes, W.~D. Dorland, and T.~Tatsuno,
\newblock Physics of Plasmas {\bf 17}, 032106 (2010).

\end{thebibliography}
\end{document}